\documentclass[lettersize,journal]{IEEEtran} 

\IEEEoverridecommandlockouts                            
\usepackage[english]{babel}
\usepackage{comment}
\usepackage{amsthm}
\usepackage{amsmath,amssymb}
\usepackage{xcolor}
\usepackage{amsfonts}
\usepackage{booktabs}
\usepackage{multirow} 
\usepackage{multicol}
\usepackage{makecell}
\usepackage{cite}
\usepackage{textcomp}
\newtheorem{definition}{Definition}
\newtheorem{remark}{Remark}
\newtheorem{theorem}{Theorem}
\newtheorem{proposition}{Proposition}

\usepackage{graphics} 
\usepackage{graphicx}
\graphicspath{ {./} }
\usepackage[rightcaption]{sidecap}
\usepackage{algorithm}
\usepackage{algpseudocode}
\usepackage{float}
\usepackage{tabularx}
\usepackage{array}
\usepackage{wrapfig}
\usepackage{enumitem}
\newlist{Properties}{enumerate}{2}
\setlist[Properties]{label=Property \arabic*., itemindent=*}

\newcommand\at[2]{\left.#1\right|_{#2}}
\makeatletter
\DeclareFontFamily{U}{tipa}{}
\DeclareFontShape{U}{tipa}{m}{n}{<->tipa10}{}
\newcommand{\arc@char}{{\usefont{U}{tipa}{m}{n}\symbol{62}}}%

\newcommand{\arc}[1]{\mathpalette\arc@arc{#1}}

\newcommand{\arc@arc}[2]{%
  \sbox0{$\m@th#1#2$}%
  \vbox{
    \hbox{\resizebox{\wd0}{\height}{\arc@char}}
    \nointerlineskip
    \box0
  }%
}
\makeatother

\usepackage[utf8]{inputenc}

\title{\LARGE \bf
A Novel Convex Layers Strategy for Circular Formation in Multi-Agent Systems

}

\author{Gautam Kumar$^{1}$ and Ashwini Ratnoo$^{2}$
\thanks{$^{1}$ Gautam Kumar is a Ph.D. student in the Department of Aerospace Engineering, Indian Institute of Science, Bangalore 560012, India.
        {\tt\small gautamkumar1@iisc.ac.in}}%
\thanks{$^{2}$ Ashwini Ratnoo is an Associate Professor(AcP) in the Department of Aerospace Engineering, Indian Institute of Science, Bangalore 560012, India.
        {\tt\small ratnoo@iisc.ac.in}}%
}

\begin{document}

\maketitle
\thispagestyle{empty}
\pagestyle{empty}

\begingroup
\renewcommand\thefootnote{}
\footnotetext{%
© 2025 IEEE. Personal use of this material is permitted. Permission from IEEE must be obtained for
all other uses, in any current or future media, including reprinting/republishing this material for
advertising or promotional purposes, creating new collective works, for resale or redistribution to
servers or lists, or reuse of any copyrighted component of this work in other works.

\vspace{0.5em}
This is the author's accepted version of the article published in
\textit{IEEE Transactions on Systems, Man, and Cybernetics: Systems}.
The final published version is available at: \texttt{https://doi.org/10.1109/TSMC.2025.3648313}.}
\addtocounter{footnote}{-1}
\endgroup

\begin{abstract}

This article considers the problem of conflict-free distribution of point-sized agents on a circular periphery encompassing all agents. The two key elements of the proposed policy include the construction of a set of convex layers (nested convex polygons) using the initial positions of the agents, and a novel search space region for each of the agents. The search space for an agent on a convex layer is defined as the region enclosed between the lines passing through the agent's position and normal to its supporting edges. Guaranteeing collision-free paths, a goal assignment policy designates a unique goal position within the search space of an agent at the initial time itself, requiring no further computation thereafter. In contrast to the existing literature, this work presents a one-shot, collision-free solution to the circular distribution problem by utilizing only the initial positions of the agents. Illustrative examples and extensive Monte-Carlo studies considering various practical attributes demonstrate the effectiveness of the proposed method.


\end{abstract}


\section{Introduction}

Swarm robotics and intelligence have garnered a lot of attention over the past few decades. This is primarily due to the growing advances in robotics and related fields like microelectronics and communication technology. In contrast to a single robot, swarms offer advantages in terms of cost, mobility, reliability, and ability to cover large areas. Applications like surveillance \cite{surveillance_tsmc}, search and rescue \cite{wang2024branch}, payload transport \cite{rao2023integrated}, and area coverage \cite{area_cover_tsmc} desire the agents in a swarm to be spatially arranged in geometric patterns like line, wedge, circle, or polygon. 

Circular formation of agents finds specific relevance in applications like target encirclement \cite{QingshanTarget2024}, ocean sampling \cite{10}, and boundary monitoring \cite{song2018circle}. In \cite{12,flocchini2017distributed}, the proposed algorithms show that it is always possible to bring a finite number of agents arbitrarily positioned in a plane to a circular formation. A two-stage policy proposed in \cite{yang2023polygon,pang2023multi} emphasizes circular formation as an intermediate configuration that can be used to eventually achieve other geometric patterns like convex and concave polygons. Besides, the distribution of multiple agents on a circular boundary offers several advantages. The work in \cite{freitas2021effects} shows that the desired spatial and temporal separation between mobile agents on a circular boundary is useful in numerous applications, such as data collections, patrolling and satellite constellations. The tracking problem in \cite{bhowmick2016tracking,brinon2014cooperative} and the target enclosing problem in \cite{yu2019cooperative} involves multiple agents forming a circular pattern around the target, which offers adaptability, robustness, complete coverage and provable performance guarantees to the multi-agent system.

Circular formation can be achieved by assigning unique goal positions for all agents on the circular boundary and then finding non-conflicting paths for the agents to move to their respective goal positions. A simple strategy in \cite{Huang2019Convhull} assigns goal positions to multiple agents along the radial direction, and then the agents move along the path connecting their initial positions and goal positions. That approach, however, fails to offer conflict-free goal assignment for agents lying on the same radial line. Another radial goal assignment policy is considered by \cite{shan2017asynchornous} wherein the agents use \textit{Sense-Process-Act} cycles at each time step and switch their goal positions if a collision with another agent is detected. In conjunction with the radial goal assignment policy, an artificial potential function-based method is proposed in \cite{yang2023polygon} to avoid collisions between agents. In \cite{alonso2011multi}, the velocity obstacle method is used to avoid inter-agent collisions as agents move to occupy predefined goal positions on a circular boundary. In \cite{katreniak2005biangular}, the circular formation strategy requires the agent closest to the circle to move along the radial line toward the circumference of the circle, while the other agents positioned on the same radial line remain stationary temporarily.

In \cite{flocchini2017distributed,di2020gathering,feletti2018uniform,das2020forming,Adhikary2021grid}, the circle formation methods essentially consider \textit{Look-Compute-Move} (LCM) cycle for realizing collision avoidance among agents. Therein, the agent's speed is commanded to be zero if a collision is detected; otherwise, the agents use a positive velocity. Further, monitoring of the agent's configuration is required at each cycle. A circle formation strategy is proposed in \cite{defago2002circle} where Voronoi diagram is constructed using the initial positions of the agents as generators. The vertex of the agent's Voronoi cell which is closest to the circle is selected as its intermediate goal point. In that approach, the non-conflicting intermediate goal assignment relies on the unboundedness of the Voronoi cells, which may not be guaranteed as the number of agents increases. The concurrent goal assignment solution proposed in \cite{turpin2014capt} requires re-evaluation when any two agents' trajectories are found conflicting at a time step. While assigning goal positions to the agents on a circular boundary in \cite{14}, the intersecting paths are considered as conflicts without assessing the temporal aspect of the collision possibilities. In all of the aforementioned works, partial or complete knowledge of the other agents' positions is required at all times or at discrete time steps. This is necessary to compute input commands of the agents such that there is no inter-agent collision while they move to occupy their respective goal positions on the circle.

The motivation for our work is to come up with a strategy that uses only the initial position of the agents and computes, at the initial time itself, a conflict-free goal assignment on the circular periphery. To the best of the authors' knowledge, none of the existing circular distribution works offer a one-shot, conflict-free goal assignment policy. This paper presents a convex layer-based approach for driving a swarm of point-sized agents on a circular boundary, which offers several key advantages. First, unlike the radial goal assignment-based circular formation strategies in \cite{yang2023polygon,shan2017asynchornous,katreniak2005biangular,alonso2011multi}, our proposed solution does not require inter-agent sensing during execution, and thereby resulting in reduced computational overload. Second, whereas the intermediate goal assignment using Voronoi-based method in \cite{defago2002circle} may not scale as the number of agents increases, the proposed goal assignment is deterministic from the outset and is scalable for a large swarm. Third, rather than relying on adjusting speeds while execution as in the LCM cycle methods \cite{flocchini2017distributed,di2020gathering,feletti2018uniform,das2020forming,Adhikary2021grid}, the proposed solution offers a precomputed and deterministic goal assignment in closed-form, which guarantees conflict-free trajectories for the agents. To summarise, the proposed solution eliminates the need for runtime replanning, which makes it suitable for agents having limited sensing and communication capabilities. The main contributions of this paper are as follows.

\begin{enumerate}
    \item A novel angular region, called the search space, is defined for each agent in the swarm. Within this search space, a goal position is defined on the circumference of a circle that encompasses all agents.

    \item By virtue of the proposed angular region and the convex layer on which an agent is located, a guarantee is deduced which rules out any collision possibility among agents. Once the goal positions are assigned, the agents move directly toward their goal position along a straight line with a prescribed speed.


    \item The proposed policy generates one-shot conflict-free trajectories deduced for any number of point-sized agents in the swarm with arbitrary initial configuration within an encompassing circle.

    \item Simulation Results demonstrate the effectiveness and scalability of the proposed method in terms of computational load and under various practical constraints that include size and dynamics of the agents, uncertainty in position measurements and communication delay.
    
\end{enumerate}

The remainder of the article is organized as follows: Section \ref{sec:2} contains the preliminaries necessary throughout the paper. The problem is formulated in Section \ref{sec:3} and the main results are presented in Section \ref{sec:4}. Simulation studies demonstrating the proposed policy are presented in Section \ref{sec:5} followed by concluding remarks in Section \ref{sec:6}.


\section{Preliminaries}\label{sec:2}

\subsection{Convex Hull}
The convex hull for a set $P$ of $n$ points, $\text{Conv}(P)$ is defined as the set of all points $p \in \mathbb{R}^2 $ such that
\begin{equation}
   p = \sum_{i=1}^n \lambda_ip_i = \lambda_1p_1+\lambda_2 p_2 +\dotsc +\lambda_n p_n,
\end{equation}
where $p_i \in P,~\lambda_i \geq 0 \in \mathbb{R},~ \forall  i=1,2,\dotsc, n$, and $\sum\limits_{i=1}^n \lambda_i = 1$. Since \textit{Graham's scan} offers a complexity of the order $\mathbb{O}(n\log{n})$ \cite{graham1972efficient}, it is used to generate Conv$(P)$ in this work.

\begin{definition}[\textnormal{\cite{36}}]
A point $V \in$ Conv$(P)$ is defined as the vertex of Conv$(P)$ if it cannot be expressed in the form of the convex combination of any two distinct points in Conv$(P)$, that is, 
\begin{equation}\label{vertex}
    V\neq c V_1 + (1-c) V_2,~ c \in[0,1] ,  
\end{equation}
where $V_1,V_2\in$ Conv$(P)$ and $V_1 \neq V_2$.\\
\end{definition}

\begin{definition}
The supporting edges of a vertex $V$ are the edges of Conv$(P)$ that intersect at $V$.
\end{definition}

\begin{definition}\label{def:ss}
The search space for a vertex $V$ of Conv$(P)$, $SS(V)$ is proposed as the angular region $[\alpha^o,\alpha^f]$ enclosed by the normals drawn at the supporting edges at $V$ (Fig. \ref{fig:search space}a). The search space range $ \Delta \alpha= \alpha^f - \alpha^o$. As an example, Fig. \ref{fig:search space}b depicts the search spaces for each of the vertices of the convex hull defined for a set of five noncollinear points. In the scenario where the points in $P$ are collinear on a line $\mathcal{L}_c$, the search space region of $p_i,~i=\{1,f\}$ is the half-plane $\Omega_k$ determined by the line $\perp \mathcal{L}_c$ and passing through $p_i$ such that $\Omega_k \cap \mathcal{L}_c/p_i=\emptyset$ (Fig. \ref{fig:search space}c). For the intermediate points $p_i,~(2\leq i\leq f-1)$ on $\mathcal{L}_c$, the search space region is the straight line $\perp \mathcal{L}_c$ and passing through $p_i$. When there is only one point in $P$, the search space region spans the entire angular space, that is, $[0,2\pi)$.
\begin{figure}[!hbt]
    \centering
    \includegraphics[width=1\columnwidth]{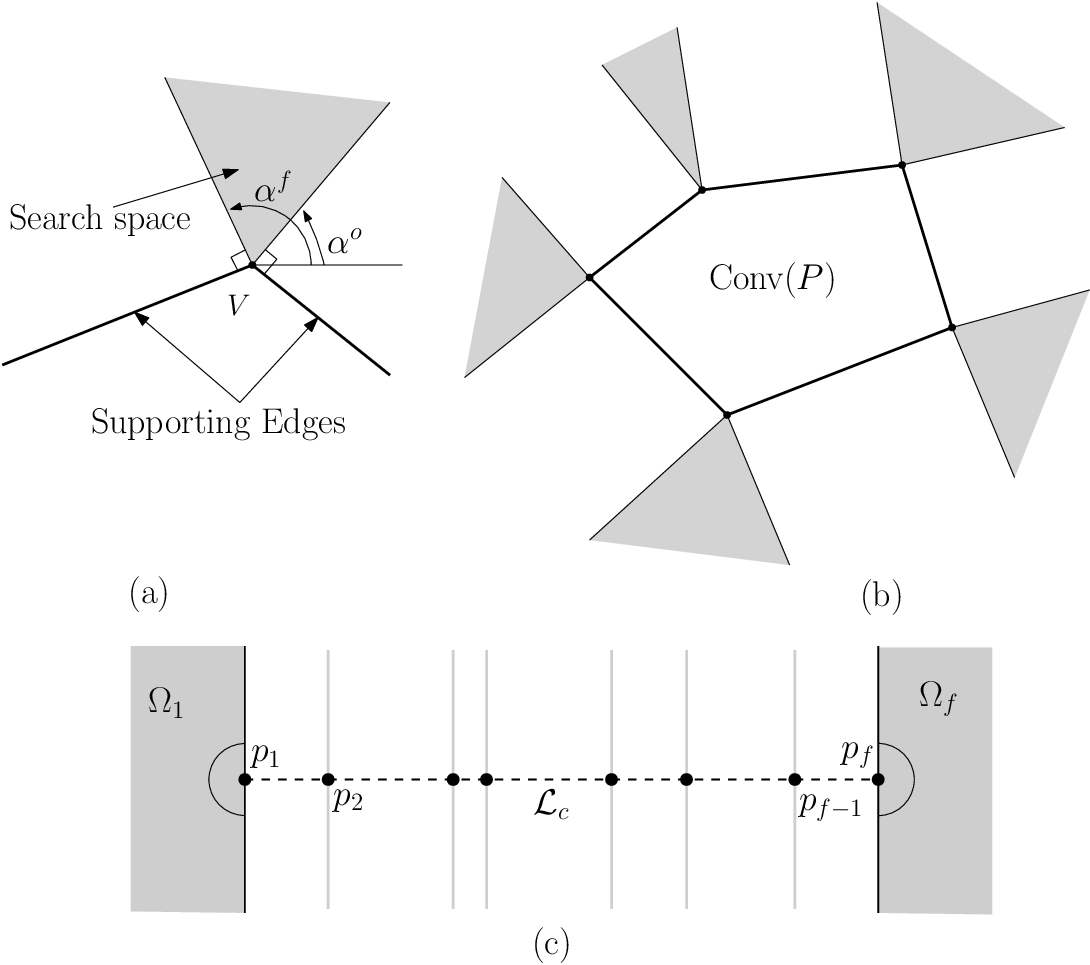}
    \caption{Search space (shaded regions): (a) for a single vertex, (b) for all vertices of a convex layer, (c) for collinear points.}
    \label{fig:search space}
\end{figure}
\end{definition}

\subsection{Convex Layers}
\begin{figure}[!hbt]
    \centering
    \includegraphics[width=\columnwidth]{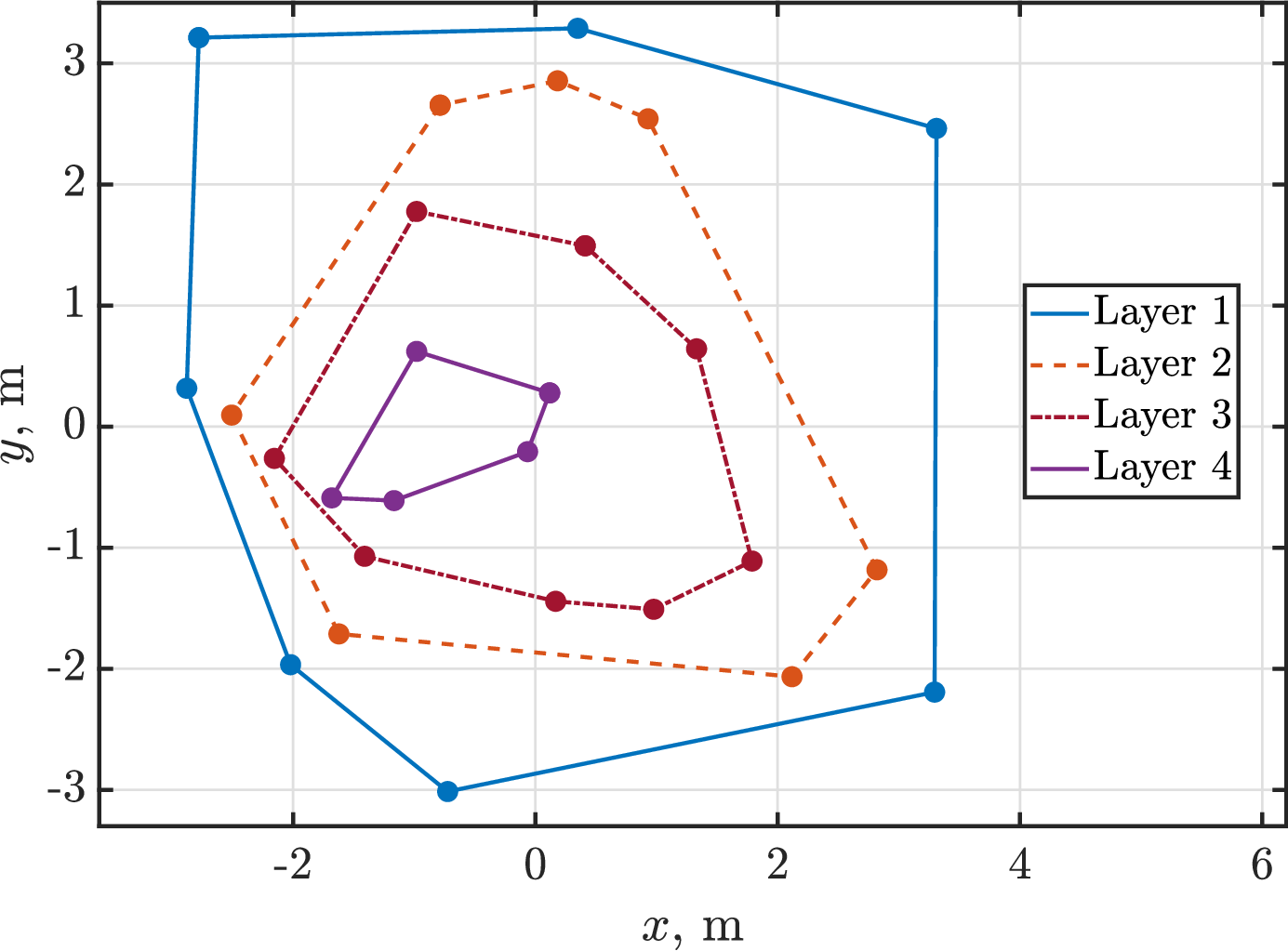}
    \caption{Convex layers for a set of 26 randomly selected points in a rectangular region bounded by the lines $y=-3.5,y=3.5,x=-3.5$, and $x=3.5$.}
    \label{fig:convex_layers}
\end{figure}

In \cite{29}, the convex layers for a set $S$ of $n$ points are defined as the set of nested convex polygons formed using the following procedure: form the convex hull of the points in $S$, delete the points from $S$ that form the vertices of the convex hull and continue the process until the number of points in $S$ is less than 3. Consider a set of randomly selected 26 points in a plane such that the $x-$ and $y-$coordinates of points satisfy $x,y \in [-3.5,3.5]$. For this example, Fig. \ref{fig:convex_layers} shows the formation of four convex layers using the aforementioned procedure. Some of the important properties of convex layers are:
\begin{enumerate}[label=(P\arabic*)]
    \item The set of convex layers for a set of points is unique.
    \item Each layer is a convex polygon.
    \item No two layers share a common vertex.
    \item For any two convex layers, one of the layers completely encompasses the other.
\end{enumerate}

 The procedure for forming convex layers is formally presented in Algorithm \ref{alg:convex_layer}. Here $S = \{ \mathbf{s}_1,\mathbf{s}_2,\dotsc,\mathbf{s}_n \}$ denote the set of $n$ points where $\mathbf{s}_i \in \mathbb{R}^2~ (1 \leq i \leq n)$. The set of convex layers is denoted by $CL = \{CL_1,CL_2,\dotsc,CL_M\}$ where $CL_1$ is the outermost layer and $CL_k \cap CL_{k+1} = CL_{k+1}~(k=1,2,\dotsc, M-1)$.  
\begin{algorithm}[!hbt]
\caption{Assigning Agents on Convex Layers \cite{29}}
\label{alg:convex_layer}
\textbf{Input : $S$ }\\
\textbf{Output : $\{L_1,L_2,\dotsc,L_M\}$ }
\begin{algorithmic}[1]
\Ensure $\mathbf{s}_i\neq \mathbf{s}_j, \; \forall\; \mathbf{s}_i,\mathbf{s}_j \in S, ~(i,j \in \{1,2,\ldots,n\})$
\State $m \gets 1$
\State $L_0 \gets \emptyset$ \Comment{$L_m$ stores vertices of $CL_m$.}
\While{number of agents in $(S - \sum_{j=0}^{m-1} L_j)> 2$}
    \If{number of agents in $(S - \sum_{j=1}^{m-1} L_j)\leq 2$ or area of Conv$(S - \sum_{j=1}^{m-1} L_j)$ == 0}
        \State $L_{M} \gets S - \sum_{j=1}^{m-1} L_j$
        \State break from the loop
    \EndIf 
 \State $L_m$ $\gets$ vertices of Conv$(S - \sum_{j=1}^{m-1} L_j)$ 
    \State $m \gets m+1$
\EndWhile
\end{algorithmic}
\end{algorithm} 

\begin{remark}\label{remark:col_case}
A trivial case may arise at the $m$th iteration in Algorithm \ref{alg:convex_layer} when the remaining points ($>2$) are found to be collinear, that is, $|(\mathbf{s}_a - \mathbf{s}_b) \times (\mathbf{s}_b - \mathbf{s}_c)| = 0, \; \forall \; \mathbf{s}_a, \mathbf{s}_b , \mathbf{s}_c \in (S-\sum_{j=1}^{m-1}L_j)  \text{ and }  a \neq b, b \neq c, a \neq c $. The algorithm ends at that iteration (see Step 4 of Algorithm \ref{alg:convex_layer}) with the remaining collinear points stored in the set $L_M~(M=m)$.
\end{remark}


\section{Circular Distribution Problem}\label{sec:3}

\begin{figure}[!hbt]
    \centering
    \includegraphics[width=0.85\linewidth]{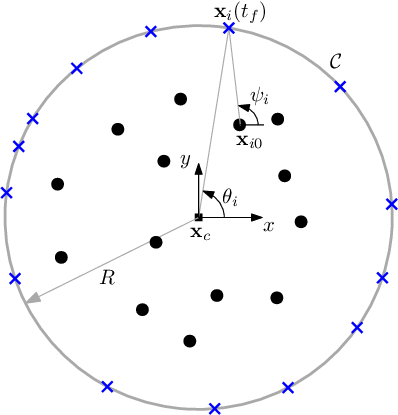}
    \caption{A sample circular distribution problem with 14 agents (circular and x-shaped markers represent agents' initial and representative goal positions, respectively).}
    \label{fig:prob_form}
\end{figure}

Consider a planar region consisting of a swarm of $n (\geq 3)$ point-sized agents. The kinematics of the $i$th agent is governed by
\begin{align}\label{eq:current_rel1}
  \begin{aligned}
    \Dot{\mathbf{x}}_i(t) & =v[\cos{\psi_i},\sin{\psi_i}], ~ \forall i = 1,2,\ldots, n.
  \end{aligned}
\end{align}
Here, $\mathbf{x}_i(t)\in \mathbb{R}^2,~ v \in \mathbb{R}^+$ and $\psi_i \in [0,2\pi)$ represent the position, the constant forward velocity, and the heading angle input, respectively, of the $i$th agent. Let $\mathcal{C}(\mathbf{x}_c,R)$ denote a circle where $\mathbf{x}_c \in \mathbb{R}^2 $ and $R>0$ are its centre and radius, respectively, and the initial positions of the agents satisfy 
\begin{align}
   &\mid \mid \mathbf{x}_{i0} - \mathbf{x}_c \mid \mid < R, &\forall i = 1,2,\dotsc,n, 
\end{align}
where $\mathbf{x}_{i0}$ is the initial position of the $i$th agent. The objective here is to determine $\psi_i$ $(i= 1,2,\dotsc,n)$ such that at some finite time $t_f^i > 0$ and in a collision-free manner, the $i$th agent occupies a unique goal position on the circumference of $\mathcal{C}$, that is,  
\begin{align}
    \mathbf{x}_i(t_f^i)&=\mathbf{x}_c+R[\cos {\theta_i},\sin {\theta_i}], ~\theta_i \neq \theta_j, ~ \forall i \neq j \label{eq:problem_goal1}\\
  \text{and, }  \mathbf{x}_i(t)&\neq \mathbf{x}_j(t),~ \forall i \neq j,0<t\leq\max(t_f^i,t_f^j).\label{eq:problem_goal2}
\end{align}
In \eqref{eq:problem_goal1} and \eqref{eq:problem_goal2}, $i,j =\{1,2,\dotsc,n\} \text{ and } \theta_i \in [0,2\pi)$ is the relative angular orientation of $\mathbf{x}_i(t_f^i)$ as measured in a fixed frame with its origin at $x_c$. Fig. \ref{fig:prob_form} shows a representative scenario of the problem. Further, this work considers the following assumptions:
\begin{enumerate}[label=(A\arabic*)]
    \item No two agents are initially collocated.
    \item Each agent is capable of moving in any direction.
    \item A centralized server has the initial position information of all agents. \label{assum:a3}
    \item The server computes and transmits heading angle input for every agent in a centralized manner.  \label{assum:a4}
    \item Owing to the fast and accurate inner loop dynamics, the low-level controllers track the prescribed $\psi_i$ and $v$ with negligible error.
\end{enumerate}

\section{Main Results}\label{sec:4}

In this section, we propose a goal assignment solution for determining a unique goal position on the circle $\mathcal{C}$ for each agent. Further, leveraging the search space for an agent using Definition \ref{def:ss}, a conflict-free strategy is devised to guide the agents towards their respective goal positions.

\subsection{Proposed Goal Assignment Policy}

\begin{figure}[!hbt]
    \centering
    \includegraphics[width =0.77 \columnwidth]{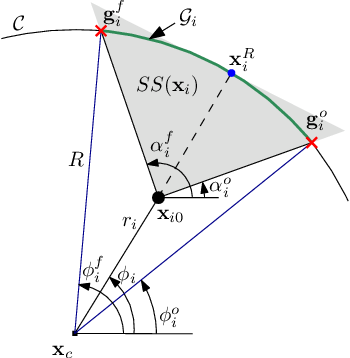}
    \caption{Goal assignment for $i$th agent.}
    \label{fig:goal_assign}
\end{figure}

Using Algorithm \ref{alg:convex_layer}, the set $CL$ of convex layers is formed using the initial positions of the agents such that the set $L_m$ ($1\leq m \leq M$) stores the vertices of $CL_m$. Accordingly, each vertex of $CL_m$ represents the initial position of an agent. Using Definition \ref{def:ss}, the search space $SS(\mathbf{x}_{i0}) ~(1\leq i \leq n)$ is constructed for the $i$th agent. Let $\mathcal{C}_b$ denotes the set of all points on $\mathcal{C}$. The set of potential goal positions, $\mathcal{G}_i$ for the $i$th agent is obtained from the intersection of $SS(\mathbf{x}_{i0})$ and $\mathcal{C}_b$, that is, 
\begin{equation}\label{eq:goal_pos_All}
    \mathcal{G}_i =\mathcal{C}_b \cap SS(\mathbf{x}_{i0}).
\end{equation}
The end points of the arc $\mathcal{G}_i$ are $\mathbf{g}_i^o$ and $\mathbf{g}_i^f$, and let $\phi_i^o$ and $\phi_i^f$ denote the polar angles of $\mathbf{g}_i^o$ and $\mathbf{g}_i^f$, respectively, in the fixed frame centered at $\mathbf{x}_c$. In Fig. \ref{fig:goal_assign}, the gray-shaded region and the green arc $\arc{\mathbf{g}_i^o \mathbf{g}_i^f}$ represent $SS(\mathbf{x}_{i0})$ and $\mathcal{G}_i$ of the $i$th agent, respectively. The following theorem presents a strategy to determine the $i$th agent's goal position $\mathbf{g}_i \in \mathcal{G}_i$ which offers the minimum Euclidean distance from $\mathbf{x}_{i0}$.

\begin{theorem}\label{th:optimal}
Consider the $i$th agent with its position $(r_i,\phi_i)$ as expressed in polar coordinate system centered at $\mathbf{x}_c $ (Fig. \ref{fig:goal_assign}). Let $SS(\mathbf{x}_{i0})$ intersect $\mathcal{C}_b$ to obtain the arc $\arc{\mathbf{g}_i^o \mathbf{g}_i^f}$ such that the polar coordinates of $\mathbf{g}_i^o$ and $\mathbf{g}_i^f$ relative to $\mathbf{x}_c$ are $(R,\phi_i^o)$ and $ (R,\phi_i^f)$, respectively $(\phi_i^o< \phi_i^f)$. The goal position $\mathbf{g}_i$ of the $i$th agent for which it travels the minimum Euclidean distance to $\mathcal{G}_i$ is  

\begin{align}\label{eq:short_path}
    \begin{aligned}
        \mathbf{g}_i = \begin{cases}
            \mathbf{x}_i^R, & \text{if }\phi_i \in [\phi_i^o,\phi_i^f] \\
            \mathbf{g}_i^o, & \text{if }\phi_i \notin [\phi_i^o,\phi_i^f] \text{ and } \mid \phi_i^o-\phi_i \mid \leq \mid \phi_i^f-\phi_i \mid\\
            \mathbf{g}_i^f,& \text{if }\phi_i \notin [\phi_i^o,\phi_i^f] \text{ and } \mid \phi_i^o-\phi_i \mid > \mid \phi_i^f-\phi_i \mid
        \end{cases}
    \end{aligned}
\end{align}
where $\mathbf{x}_i^R = \mathbf{x}_c +R[\cos{\phi_i},\sin{\phi_i}]$ is the point on $\mathcal{C}_b$ that corresponds to the shortest path to $\mathcal{C}_b$ for the $i$th agent.
\end{theorem}

\begin{proof}
The distance between the $i$th agent and a point $p(R,\phi) \in  \mathcal{C}_b$ can be expressed as:
\begin{align}
\begin{aligned}
    D_i(\phi)=\sqrt{ R^2+r_i^2-2Rr_i\cos (\phi-\phi_i)}.
\end{aligned}
\end{align}
The objective is to find $\phi$ which minimizes $ D_i(\phi)$, that is, 
\begin{align}\label{eq:minimze_D}
      &\min_\phi D_i(\phi),&
      \text{ subject to}&~~ \phi_i^o-\phi \leq 0, ~\phi - \phi_i^f\leq 0 .
\end{align}
The Lagrangian Multiplier method is used to solve the constrained optimization problem in (\ref{eq:minimze_D}). Accordingly, the Lagrangian of (\ref{eq:minimze_D}) is expressed as
\begin{align}\label{eq:lagrangian}
\begin{aligned}
    \mathcal{L}(\phi,\mu_1,\mu_2)&=\sqrt{ R^2+r_i^2-2Rr_i\cos (\phi-\phi_i)}\\& +\mu_1(\phi -\phi_i^f)+\mu_2(\phi_i^o-\phi),
    \end{aligned}
\end{align}
where $\mu_1,\mu_2\geq 0$ are Lagrange multipliers. Let $\mathcal{D}_{\phi\phi} = \dfrac{\partial^2D_i}{\partial \phi^2} =\dfrac{Rr_i\cos (\phi-\phi_i)}{D_i} - \left(\dfrac{Rr_i\sin (\phi-\phi_i)}{D_i^{3/2}}\right)^2$. Further, different combinations of active constraints are analyzed for $\mathcal{L}$ and the feasibility of solutions is checked.

\textit{Case 1}: $\mu_1=\mu_2=0 $. In this case, the Lagrangian in (\ref{eq:lagrangian}) reduces to $\mathcal{L} =D_i$, and the gradient and the Hessian of $\mathcal{L}$ are
\begin{align}
    \nabla \mathcal{L} &=   \dfrac{Rr_i\sin (\phi-\phi_i)}{D_i}, \label{eq:grad_case1}\\
    \nabla^2 \mathcal{L} &= \mathcal{D}_{\phi\phi}.\label{eq:hess_case1}
\end{align}
In accordance with the first-order necessary condition, $\nabla \mathcal{L}=0$ is evaluated using \eqref{eq:grad_case1} to obtain the critical points of $\mathcal{L}$.
\begin{align}\label{eq:first_deriv_minima}
\begin{aligned}
   \nabla \mathcal{L} &= \dfrac{Rr_i}{D_i}\sin (\phi^*-\phi_i)=0\\
    \implies\phi^*&=\{\phi_i,\pi +\phi_i\}.
    \end{aligned}
\end{align}
To determine the local minimum point from the critical points $\phi^*=\{\phi_i,\pi +\phi_i\}$ obtained in (\ref{eq:first_deriv_minima}), the second-order necessary condition is checked by evaluating $\nabla^2\mathcal{L}$ in (\ref{eq:hess_case1}) at $\phi^*$, that is,
 \begin{align}
         \at{\nabla^2 \mathcal{L}}{\phi^* =\phi_i}&= \dfrac{Rr_i}{D_i}>0 \label{eq:case1_min}, \\
         \at{\nabla^2 \mathcal{L}}{\phi^* =\phi_i+\pi}&=- \dfrac{Rr_i}{D_i}<0.
\end{align}
Since $ \at{\nabla^2 \mathcal{L}}{\phi^* =\phi_i}>0 $ from \eqref{eq:case1_min}, the solution that minimizes $D_i$ is $\phi^*=\phi_i$. Given $\mu_1=\mu_2=0$, the solution $\phi^*=\phi_i$ is feasible when
\begin{align}\label{eq:goal_case1}
   & \phi_i^o \leq\phi^* =\phi_i \leq \phi_i^f. \\
   \text{Here, }&\phi^*=\phi_i \in [\phi_i^o,\phi_i^f]\implies \mathbf{g}_i=\mathbf{x}_i^R.
\end{align}

\textit{Case 2}: $\mu_1=0,\mu_2>0$. Here, using \eqref{eq:lagrangian}, the Lagrangian is obtained as $\mathcal{L} =D_i+\mu_2(\phi_i^o - \phi)$, and the gradient and the Hessian of $\mathcal{L}$ are
\begin{align}
    \nabla \mathcal{L} &=   \begin{bmatrix}
        \dfrac{Rr_i\sin (\phi-\phi_i)}{D_i} -\mu_2\\
        (\phi_i^o - \phi)
    \end{bmatrix}, \label{eq:grad_case2} \\  \nabla^2 \mathcal{L} &= \begin{bmatrix}
        \mathcal{D}_{\phi\phi} & -1\\
        -1 &0
    \end{bmatrix} .\label{eq:hess_case2}
\end{align}
Using \eqref{eq:grad_case2} and applying the first-order necessary condition on $\mathcal{L}$ to find its critical points,
\begin{align}
   \nabla \mathcal{L} &= \begin{bmatrix}
        \dfrac{Rr_i\sin (\phi-\phi_i)}{D_i} -\mu_2\\
        (\phi_i^o - \phi)
    \end{bmatrix}=0,\\
    \implies \phi^*&=\phi_i^o, ~\mu_2^* =\dfrac{Rr_i\sin (\phi_i^o-\phi_i)}{D_i} . \label{eq:mu2} 
\end{align}

To check the second-order necessary condition, $\nabla^2 \mathcal{L}$ in (\ref{eq:hess_case2}) is evaluated at the point $(\phi^o_i,\mu_2^*)$. 
\begin{equation}
    \at{\nabla^2\mathcal{L}}{\phi=\phi^o_i,\mu_2=\mu_2^*}=\begin{bmatrix}
       \at{ \mathcal{D}_{\phi\phi}}{\phi=\phi^o_i} & -1\\
       -1 &0
    \end{bmatrix}.
\end{equation}

Let $a = \at{ \mathcal{D}_{\phi\phi}}{\phi=\phi^o_i}$. Then, the eigenvalues of $\at{\nabla^2\mathcal{L}}{\phi=\phi^o_i,\mu_2=\mu_2^*}$ are

\begin{equation}\label{eq:eig1}
    \lambda_{1,2} = \dfrac{a \pm \sqrt{a^2+4}}{2} .
\end{equation}
In (\ref{eq:eig1}), $\forall a \in \mathbb{R}$, the eigenvalues are mixed $\implies \at{\nabla^2\mathcal{L}}{\phi=\phi^o_i,\mu_2=\mu_2^*}$ is indefinite and $(\phi^o_i,\mu_2^*)$ is a saddle point.

\textit{Case 3}: $\mu_1>0,\mu_2=0$. In this case, the Lagrangian in (\ref{eq:lagrangian}) is given by $\mathcal{L} =D_i+\mu_1(\phi -\phi_i^f)$, and the gradient and the Hessian of $\mathcal{L}$ are
\begin{align}
    \nabla \mathcal{L} &=   \begin{bmatrix}
        \dfrac{Rr_i\sin (\phi-\phi_i)}{D_i} +\mu_1\\
        (\phi_i^o - \phi)
    \end{bmatrix},\label{eq:grad_case3}\\
    \nabla^2 \mathcal{L}& = \begin{bmatrix}
        \mathcal{D}_{\phi\phi} & 1\\
        1 &0
    \end{bmatrix} \label{eq:hess_case3}
\end{align}
Following the first-order necessary condition, $\nabla \mathcal{L} = 0$ using \eqref{eq:grad_case3},
\begin{align}
   \nabla \mathcal{L} &= \begin{bmatrix}
        \dfrac{Rr_i\sin (\phi-\phi_i)}{D_i} +\mu_1\\
        ( \phi-\phi_i^f)
    \end{bmatrix}=0,\\
    \implies \phi^*&=\phi_i^f, ~\mu_1^* =-\dfrac{Rr_i\sin (\phi_i^f-\phi_i)}{D_i} . \label{eq:mu1}
\end{align}
For the second-order necessary condition, $\nabla^2 \mathcal{L}$ in \eqref{eq:hess_case3} is evaluated at $\phi=\phi^f_i,\mu_1=\mu_1^*$ as 
\begin{equation}
    \at{\nabla^2\mathcal{L}}{\phi=\phi^f_i,\mu_1=\mu_1^*}=\begin{bmatrix}
       \at{ \mathcal{D}_{\phi\phi}}{\phi=\phi^f_i} & 1\\
       1 &0
    \end{bmatrix}.
\end{equation}
Let $a = \at{ \mathcal{D}_{\phi\phi}}{\phi=\phi^f_i}$. Accordingly, the eigenvalues of $\at{\nabla^2\mathcal{L}}{\phi=\phi^f_i,\mu_1=\mu_1^*}$ are

\begin{equation}\label{eq:eig2}
    \lambda_{1,2} = \dfrac{a \pm \sqrt{a^2+4}}{2} .
\end{equation}
In (\ref{eq:eig2}), $\forall a \in \mathbb{R}$, the eigenvalues are mixed $\implies \at{\nabla^2\mathcal{L}}{\phi=\phi^f_i,\mu_1=\mu_1^*}$ is indefinite and $(\phi^f_i,\mu_1^*)$ is a saddle point.

Two saddle points, $(\phi_i^o,\mu_2^*)$ and $(\phi_i^f,\mu_1^*)$, are obtained from Cases 2 and 3, respectively. From \eqref{eq:mu1},  $\mu_1^*>0 \implies \pi+\phi_i^f>\phi_i>\phi_i^f$ and from \eqref{eq:mu2},  $\mu_2^*>0 \implies \phi_i^o-\pi<\phi_i<\phi_i^o$. Combining both these cases, the two saddle points are now analyzed to find $\mathbf{g}_i$ when $\phi_i \notin [\phi_i^o,\phi_i^f] $.
\begin{align}\label{eq:compare}
    \begin{aligned}
        D_i(\phi_i^o) &= \mid \mid \mathbf{g}_i^o -\mathbf{x}_{i0}\mid \mid= \sqrt{ R^2+r_i^2-2Rr_i\cos (\phi_i^o-\phi_i)},\\
        D_i(\phi_i^f) &=\mid \mid \mathbf{g}_i^f -\mathbf{x}_{i0}\mid \mid= \sqrt{ R^2+r_i^2-2Rr_i\cos (\phi_i^f-\phi_i)}.
    \end{aligned}
\end{align}
Comparing $D_i(\phi_i^o)$ and $D_i(\phi_i^f)$ from \eqref{eq:compare},
\begin{align}
\label{eq:phi_o}       \mid \phi_i^o-\phi_i \mid \leq \mid \phi_i^f-\phi_i \mid &\implies  D_i(\phi_i^o) \leq D_i(\phi_i^f),\\
 \label{eq:phi_f}      \mid \phi_i^o-\phi_i \mid > \mid \phi_i^f-\phi_i \mid &\implies  D_i(\phi_i^o) > D_i(\phi_i^f).
\end{align}
Using \eqref{eq:phi_o} and \eqref{eq:phi_f}, $\mathbf{g}_i=\mathbf{g}_i^o$ when $\mid \phi_i^o-\phi_i \mid ~ \leq ~ \mid\phi_i^f-\phi_i \mid$ and $\mathbf{g}_i=\mathbf{g}_i^f$ when $\mid \phi_i^o-\phi_i \mid > \mid \phi_i^f-\phi_i \mid$. Since $\mathbf{g}_i^o$ and $\mathbf{g}_i^f$ correspond to saddle points of $\mathcal{L}$ and may not necessarily coincide with $\mathbf{x}_i^R$, the resulting assignment in Cases 2 and 3 may not, in general, correspond to the shortest path from $\mathbf{x}_{i0}$ to $\mathcal{C}_b$.

\textit{Case 4}: $\mu_1>0,\mu_2>0$. This case is infeasible as $\phi_i^o< \phi_i^f$ and both inequality constraints in (\ref{eq:minimze_D}) cannot be satisfied simultaneously.\end{proof}

\begin{figure}[!hbt]
    \centering
    \includegraphics[width=0.8\columnwidth]{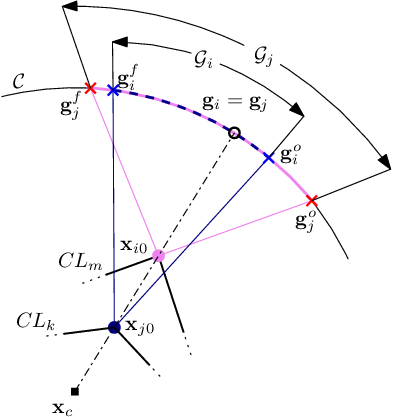}
    \caption{An example case for non-unique goal assignment (The blue and magenta arcs represent $\mathcal{G}_i$ and $\mathcal{G}_j$, respectively).}
    \label{fig:non_unique_goal}
\end{figure}

Following the goal assignment strategy discussed in Theorem \ref{th:optimal}, the next challenge is to ensure that each agent is assigned a unique goal. Although the policy proposed in (\ref{eq:short_path}) designates the goal position for each agent on the circumference of $\mathcal{C}$, it does not ensure a unique goal assignment for certain initial positional arrangements of the agents. An example of such a configuration is shown in Fig. \ref{fig:non_unique_goal} where $\mathcal{G}_i \subset \mathcal{G}_j $ and $\phi_i=\phi_j$. Here, the goal positions for agents $i$ and $j$ are found to be collocated using the policy in (\ref{eq:short_path}).

\begin{figure}[!hbt]
    \centering
    \includegraphics[width=0.95\columnwidth]{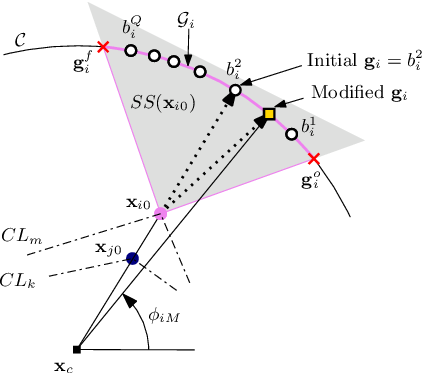}
    \caption{Demonstration of Algorithm 2 for solving conflicting goal assignment for agents $i$ and $j$.}
    \label{fig:algo1}
\end{figure}

To rule out the possibility of conflicting goal assignment, Algorithm \ref{alg:goal_assign} is proposed which assigns a unique goal for each of the agents irrespective of their initial positional arrangement. Algorithm 2 assigns goal positions in sequence, starting from the agents corresponding to $L_M$, followed by $L_{M-1}$, $L_{M-2}$, and continuing until $L_1$. Therein, $\mathcal{P}$ is the set of goal positions assigned in previous iterations, and $\mathcal{B}_i$ is the set of points in $\mathcal{P}$ that lie on $\arc{\mathbf{g}_i^o \mathbf{g}_i^f}$, that is, 
\begin{align}\label{eq:conf_goalpos}
    \mathcal{B}_i&=\mathcal{P}\cap \mathcal{G}_i= \{b_i^1,b_i^2,\dotsc,b_i^Q \},
\end{align}
where $Q$ is a non-negative integer, that is, $Q \in \mathbb{Z}_{\geq 0}$. In \eqref{eq:conf_goalpos}, if $\mathcal{P}\cap \mathcal{G}_i=\emptyset$, then $\mathcal{B}_i = \emptyset$ and $Q =0$. Points in $\mathcal{B}_i$ are assumed to be numbered counterclockwise around $\mathbf{x}_c$.  Consider the set $\Phi_i$ where each element $\Phi_i^k ~(1\leq k \leq Q+2)$ is defined by the angular position of the elements in the set $\{ \mathbf{g}_i^o,b_i^1,b_i^2,\ldots,b_i^Q, \mathbf{g}_i^f \}$ with
\begin{align}\label{eq:PHI}
    \begin{aligned}
        \Phi_i^1 \leq \Phi_i^2<  \ldots< \Phi_i^{Q+1} \leq \Phi_i^{Q+2}. 
    \end{aligned}
\end{align}
A goal conflict for the $i$th agent occurs if
\begin{align}\label{eq:prev_assign_goal}
         \mathcal{B}_i \cap \mathbf{g}_i&= \mathbf{g}_i=b_i^q,& \text {where }q \in \mathbb{Z}^+, 1\leq q\leq Q.
\end{align}
The goal position $\mathbf{g}_i$ in \eqref{eq:prev_assign_goal} must be modified to ensure a non-unique goal assignment. For the $i$th agent, the angular position $\phi_{iM}$ of unallocated goal positions within $\mathcal{G}_i$ satisfy
\begin{equation}\label{eq:range_phiM}
    \phi_{iM} \in [\Phi_i^1,\Phi_i^{Q+2}], ~ \phi_{iM} \neq \Phi_i^k,~\forall k = 2,3,\ldots,Q+1.
\end{equation}
From all possible values in \eqref{eq:range_phiM}, the objective is to select $\phi_{iM}$ such that the modified goal position is close to $\mathbf{g}_i$ governed by Theorem \ref{th:optimal}. Using \eqref{eq:PHI} and (\ref{eq:prev_assign_goal}), the corresponding angular position of $b_i^q=\mathbf{g}_i$ relative to $\mathbf{x}_c$ corresponds to $(q+1)$th element in $\Phi_i$, that is, $\Phi_i^{q+1}$. Accordingly, $\mathbf{g}_i$, in \eqref{eq:prev_assign_goal}, is recomputed as follows: 
\begin{align}
     \mathbf{g}_i &= \mathbf{x}_c +R 
            [\cos{\phi_{iM}},\sin{\phi_{iM}}],  \label{eq:mod_goal}\\
        \phi_{iM} &=  \begin{cases}
            (1-\delta)\Phi_i^{q+1} +\delta\Phi_i^{q}, \\ \hspace{1.2cm} \text{if }  |\Phi_i^{q+1} -\Phi_i^{q}|\geq  |\Phi_i^{q+1} -\Phi_i^{q+2}|\\
            (1-\delta)\Phi_i^{q+1} +\delta\Phi_i^{q+2},\\  \hspace{1.2cm} \text{if }  |\Phi_i^{q+1} -\Phi_i^{q}|<  |\Phi_i^{q+1} -\Phi_i^{q+2}|
        \end{cases}. \label{eq:mod_goal1}
\end{align}
where $0<\delta < 1$ is a constant. From \eqref{eq:mod_goal} and \eqref{eq:mod_goal1}, the direction in which the goal position $\mathbf{g}_i$ is shifted is determined by comparing the angular separation of $b_i^q$ with its immediate neighbors, $b_i^{q-1}$ and $b_i^{q+1}$, in $\mathcal{B}_i$. Accordingly, if the separation is greater in the clockwise direction, that is, $|\Phi_i^{q+1} -\Phi_i^{q}|\geq  |\Phi_i^{q+1} -\Phi_i^{q+2}|$, the shift direction is clockwise; otherwise, it is counterclockwise. The direction of the shift is conventionally chosen clockwise when $|\Phi_i^{q+1} -\Phi_i^{q}|=  |\Phi_i^{q+1} -\Phi_i^{q+2}|$. This procedure is formally presented in Algorithm \ref{alg:goal_assign}. Consider again the example shown in Fig. \ref{fig:non_unique_goal}. Using Algorithm \ref{alg:goal_assign}, $\mathbf{g}_i$ is modified as shown in Fig. \ref{fig:algo1}. 
\begin{remark}
    The value of $\delta$ determines how close the recomputed goal is to the goal assigned using \eqref{eq:short_path}. Lower the $\delta$, lower is the angular separation between $\mathbf{g}_i$ computed through \eqref{eq:short_path} and updated using \eqref{eq:mod_goal}.
\end{remark}

\begin{algorithm}[!hbt]
\caption{Goal Assignment Policy}
\label{alg:goal_assign}
\textbf{Input : $\{\mathbf{x}_1(0),\mathbf{x}_2(0),\dotsc,\mathbf{x}_n(0)\},~\{L_1,L_2,\ldots,L_M \}$ }\\
\textbf{Output : $\{\mathbf{g}_1,\mathbf{g}_2,\ldots,\mathbf{g}_n \}$ }
\begin{algorithmic}[1]

\State $m \gets M$ \Comment{Goal assignment starts with $M$th layer}
\State $\mathcal{P}\gets \emptyset$ \Comment{Stores assigned goal positions}
\While{$m \geq 1$}
    \For{each agent $i$ in $L_m$}
    \State $\mathcal{G}_i \gets SS(\mathbf{x}_{i0}) \cap \mathcal{C}_b$.
      \State Find $ \mathbf{g}_i$ using policy (\ref{eq:short_path}).
      \State $\mathcal{B}_i \gets \mathcal{G}_i \cap \mathcal{P}$.
      \If{$\mathbf{g}_i \cap \mathcal{B}_i \neq \emptyset$}
       \State Modify $\mathbf{g}_i$ using policy (\ref{eq:mod_goal}).
      \EndIf   
            \State $\mathcal{P} \gets \{\mathcal{P},\mathbf{g}_i \}$.
    \EndFor
    \State $m \gets m - 1$.
\EndWhile
\end{algorithmic}
\end{algorithm}

\begin{proposition}\label{prop:mod_goal}
    The recomputed goal position of the $i$th agent in (\ref{eq:mod_goal}) also lies in its set of potential goal positions, that is, $\mathbf{g}_i \in \mathcal{G}_i$.  
\end{proposition}
\begin{proof}
Using \eqref{eq:mod_goal},
\begin{align}\label{eq:prop2}
   \phi_{iM} \in \begin{cases}
        (\Phi_i^q ,\Phi_i^{q+1}), & \text{if  $|\Phi_i^{q+1} -\Phi_i^{q}|\geq  |\Phi_i^{q+1} -\Phi_i^{q+2}|$} \\
        (\Phi_i^{q+1} ,\Phi_i^{q+2}), & \text{if  $|\Phi_i^{q+1} -\Phi_i^{q}|<  |\Phi_i^{q+1} -\Phi_i^{q+2}|$}
    \end{cases}.
\end{align}
From the definition of the set $\Phi_i$, $ \Phi_i^1=\phi_i^o$ and $\Phi_i^{Q+2}=\phi_i^f$. From Theorem \ref{th:optimal}, the minor arc formed by the angles $\phi_i^o,\phi_i^f$ on $\mathcal{C}$ is $\mathcal{G}_i = \arc{\mathbf{g}_i^o \mathbf{g}_i^f}$. Using (\ref{eq:PHI}) and (\ref{eq:prop2}), we have 
\begin{equation}\label{eq:propos1}
    \Phi_i^1 \leq \Phi_i^q <\Phi_i^{q+1}<\Phi_i^{q+2}\leq \Phi_i^{Q+2} ~(\forall q = 1,2,\ldots Q).
\end{equation}
From \eqref{eq:propos1}, $ \phi_{iM} \in (\Phi_i^1, \Phi_i^{Q+2})$. Since the recomputed goal position $\mathbf{g}_i$ corresponds to the polar angle $\Phi_{iM}$, $\mathbf{g}_i$ lies on $\arc{\mathbf{g}_i^o \mathbf{g}_i^f}$ or $\mathbf{g}_i \in \mathcal{G}_i$.
\end{proof}

\begin{remark}
    Using Proposition \ref{prop:mod_goal}, the modified goal position of the $i$th agent lies within $SS(\mathbf{x}_{i0})$.
\end{remark}
\begin{remark}
    The heading angle input for the $i$th agent, $\psi_i$ is obtained by taking the argument of the vector $\overrightarrow{\mathbf{x}_{i0}\mathbf{g}_i}$, that is, 
\begin{equation}\label{eq:psi}
    \psi_i = \tan^{-1}\left( \dfrac{g_{iy}-y_{i0}}{g_{ix}-x_{i0}}\right),
\end{equation}
where $\mathbf{g}_i =[g_{ix},g_{iy}]$ and $\mathbf{x}_{i0}=[x_{i0},y_{i0}]$. Further, the final time $t_f^i$ is calculated by considering a straight line joining $\mathbf{x}_{i0}$ and $\mathbf{g}_i$ with agent moving at constant speed $v$, that is, 
\begin{align}
    t_f^i = \dfrac{\mid \mid \mathbf{g}_i -\mathbf{x}_{i0} \mid \mid}{v}.
\end{align}
\end{remark}
\begin{remark}
    The $i$th agent $(\forall i =1,2,\dotsc,n)$ employs a constant speed $v$ along $\psi_i$ obtained using (\ref{eq:psi}) during the interval $[0,t_f^i)$ and stops when it reaches $\mathbf{g}_i$.
\end{remark}
\subsection{Result on Guaranteed Inter-agent Collision Avoidance}\label{sec:4a}

For the goal position assigned to each agent using Algorithm \ref{alg:goal_assign}, the following theorems establish that there are no inter-agent collisions as the agents move towards their respective $\mathbf{g}_i$.

\begin{figure}[!hbt]
    \centering
    \includegraphics[width = 0.67\columnwidth]{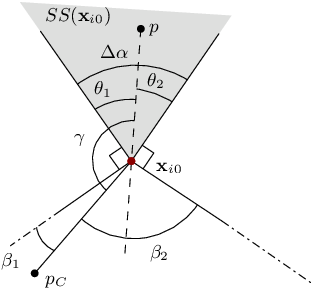}
    \caption{Collision avoidance property within $SS(\mathbf{x}_{i0})$.}
    \label{fig:dist_close}
\end{figure}

\begin{theorem}\label{th:short_path}
For the $i$th agent in $CL_m$, a point $p \in SS(\mathbf{x}_{i0})$ satisfies $\mid \mid p - \mathbf{x}_{i0} \mid \mid < \mid \mid p - p_{C} \mid \mid,$ where $p_{C} \in$ Conv$(CL_m)-\{\mathbf{x}_{i0}\}$.
\end{theorem}
\begin{proof}  In Fig. \ref{fig:dist_close}, let $\gamma$ be the included angle between the sides $p\mathbf{x}_{i0}$ and $\mathbf{x}_{i0}p_C$ of the triangle $\Delta_i$ formed by the points $\{p_C,\mathbf{x}_{i0},p\}$. As shown in Fig. \ref{fig:dist_close}, let $\theta_1,\theta_2$ be the angles formed by the segment $p\mathbf{x}_{i0}$ with the boundaries of $SS(\mathbf{x}_{i0})$, and $\beta_1,\beta_2$ be the angles formed by the segment $p_C\mathbf{x}_{i0}$ with the supporting edges of $\mathbf{x}_{i0}$. From the geometry in Fig. \ref{fig:dist_close},
\begin{align}
    0 \leq \theta_1 \leq \Delta \alpha, 0 \leq \theta_2 \leq \Delta \alpha&, \text{and} ~\theta_1+\theta_2 =\Delta \alpha \label{eq:theta_th}. \\
    0 < \beta_1 < \pi,&~ 0 < \beta_2 < \pi.  \label{eq:beta_th}
\end{align}
Here, $\gamma$ is obtained as
\begin{align}\label{eq:gamma}
    \gamma = \begin{cases}
        \theta_1+\dfrac{\pi}{2}+\beta_1,& \text{if $p_C$ is left of $\overleftrightarrow{p\mathbf{x}}_{i0}$} \vspace{0.05cm}\\ 
        \theta_2+\dfrac{\pi}{2}+\beta_2,& \text{if $p_C$ is right of $\overleftrightarrow{p\mathbf{x}}_{i0}$}\\
        \pi, & \text{if $p_C$ is on $ \overleftrightarrow{p\mathbf{x}}_{i0}$}
    \end{cases}.
\end{align}
Using (\ref{eq:theta_th})-(\ref{eq:gamma}), $\gamma\geq \pi/2$ is always the largest angle in $\Delta_i, ~\forall p_C \in CL_m, ~ \forall p \in SS(\mathbf{x}_{i0})$. Hence, $\mid \mid p - \mathbf{x}_{i0} \mid \mid < \mid \mid p - p_{C} \mid \mid$.\end{proof}

\begin{remark}\label{rem:prop_ss}
    Using Theorem \ref{th:short_path} and the goal assignment policy in Algorithm \ref{alg:goal_assign}, any point lying on the straight line segment connecting an agent's initial and goal positions is closer to itself as compared to any agent lying on the same or inner convex layers. 
\end{remark}

\begin{theorem}\label{th:final}
    Consider any two distinct agents $\mathcal{A}_i\in CL_m$ and $\mathcal{A}_j\in CL_k$ whose goal positions are $\mathbf{g}_i$ and $\mathbf{g}_j$, respectively. As both agents move with the identical prescribed speed $v$ on the straight-line path connecting their initial positions to their respective goal positions, they do not collide.
\end{theorem}
\begin{proof}
Consider $\mathbf{z}_i \in \overline{\mathbf{x}_{i0}\mathbf{g}_i}$. Without loss of generality, assume $m\leq k$. From Algorithm \ref{alg:convex_layer}, $CL_k \subset CL_m\implies \mathbf{x}_{j0} \in CL_m $. Using Remark \ref{rem:prop_ss}, 
\begin{equation}\label{eq:path_ij}
    \mid \mid\mathbf{z}_i-\mathbf{x}_{i0}\mid \mid <\mid \mid\mathbf{z}_i-\mathbf{x}_{j0}\mid \mid.
\end{equation}
From (\ref{eq:path_ij}), $\mathcal{A}_i$ reaches $\mathbf{z}_i$ prior to $\mathcal{A}_j$ as both agents move with the same speed $v$. Hence  $\forall t \in [0,t_f^i]$, there is no collision between $\mathcal{A}_i$ and $\mathcal{A}_j$. For $t> t_f^i, ~\mathbf{x}_i(t) = \mathbf{g}_i$. Using the convex property of $\mathcal{C}$, the only point $\mathbf{z}_j \in \overline{\mathbf{x}_{j0}\mathbf{g}_j}$ that lies on $\mathcal{C}_b$ is $\mathbf{g}_j$. Since $\mathbf{g}_i \neq \mathbf{g}_j$ and $\mathbf{g}_i\in \mathcal{C}_b $, $\mathbf{g}_i$ does not lie on the straight path joining $\mathbf{x}_{j0}$ and $\mathbf{g}_j$, that is, $ \mathbf{g}_i \notin \overline{\mathbf{x}_{j0}\mathbf{g}_j}$. This rules out any collision possibility for $\mathcal{A}_i$ when $t> t_f^i$. 
\end{proof}
\begin{remark}
    In conjunction with the results in Theorems (\ref{th:short_path}) and (\ref{th:final}), Algorithm \ref{alg:goal_assign} ensures a one-shot, collision-free unique goal assignment on $\mathcal{C}_b$ for each of the agents. Further, the assignment uses only the initial position information of the agents.
\end{remark}

\subsection{Computational Complexity}

    In Algorithm \ref{alg:convex_layer}, the maximum possible number of iterations or convex layers for $n$ agents is $\lfloor n/3\rfloor$, and since the complexity order of \textit{Graham's Scan} is $\mathcal{O}(n\log(n))$, the overall complexity order of Algorithm \ref{alg:convex_layer} is $\mathcal{O}(n^2\log(n))$. The checks for the conflicting goal assignment in steps (7-11) in Algorithm \ref{alg:goal_assign} correspond to computational complexity of the order $\mathcal{O}(\log(n))$. Since the iteration considers $n$ agents, the complexity order of Algorithm \ref{alg:goal_assign} is $\mathcal{O}(n\log(n))$. Accordingly, the overall complexity order of the proposed approach in the worst-case scenario is $\mathcal{O}(n^2\log(n))$. Compared to that, the complexity order for concurrent goal assignment at discrete time step in \cite{turpin2014capt} is $\mathcal{O}(n^3)$, while the LCM cycle-based approach for circular formation in \cite{flocchini2017distributed} has the complexity order of $\mathcal{O}(In^2)$ where $I$ is the number of iterations performed for detecting and resolving conflict along the path. In \cite{flocchini2017distributed}, with an increase in the number of agents and radius of the encompassing circle, the increase in $I$ is substantial and the LCM cycle-based approach is likely to face a significantly higher computational burden. 
    
 \begin{remark}   
    The procedure of convex layer construction requires repeated convex hull computations of remaining points across the iterations. The convex hull computation for the $m$th convex layer can be done in $\mathcal{O}(k\log k)$, where $k$ is the number of remaining points in that iteration. Summing over all the layers gives 
    \begin{align}
        \sum_{i=1}^M O\left(k_i \log k_i\right) \leq O\left(n^2 \log n\right).
    \end{align}
    However, in practice, especially with random initial positions of the agents, the majority of points are eliminated in the early iterations, and the number of remaining points reduces drastically with an increase in iterations in Algorithm \ref{alg:convex_layer}. This leads to the overall observed runtime of the proposed approach close to linear or log-linear growth ($\mathcal{O}(n\log n)$). The entire policy, using only the initial positions of the agents, is executed at $t=0$. This makes the proposed idea highly applicable as a pre-mission planning method for robotic swarms. Hence, the proposed policy not only offers a conflict-free solution at the initial time but also offers computational advantages as the problem scales.
\end{remark}

\subsection{Discussion}

The inter-agent collision avoidance analysis in Section \ref{sec:4a} involves two underlying assumptions: (a) identical speed assignment to each agent, and (b) the agents are point-sized. In this section, we discuss the implications of relaxing these assumptions and delve into practical implementation aspects of the proposed policy in real-world scenarios.

The analysis in Theorem \ref{th:final} depends both on prescribing identical speed to each agent and the goal assignment policy in Algorithm \ref{alg:goal_assign}. It is important to note here that identical speed assignment for agents is only a sufficient condition for guaranteeing collision avoidance. Besides identical speed, various speed assignment policies exist that can be shown to ensure inter-agent collision avoidance. For example, consider a speed assignment policy which assigns same speed $v_m$ to all agents on the $m$th convex layer $CL_m$ $(1\leq m\leq M)$ with $v_1>v_2>\ldots>v_M$. Following this speed assignment, it can be shown using a similar analysis as in Theorem \ref{th:final} that collision possibility between any pair of agents is ruled out. Nevertheless, identical speed assignment offers flexibility (any positive speed can be considered as the prescribed speed), particularly catering to a swarm of agents, and is hence considered in the work.

While the results in Section \ref{sec:4} present an important breakthrough in developing a one-shot solution to the circular distribution problem, considering the size of the agents is important in evaluating the performance in real-world scenarios. A case study assessing the effectiveness and scalability of the proposed policy while considering disc-shaped agents is carried out in Section \ref{sec:6d}.

From Assumptions \ref{assum:a3} and \ref{assum:a4}, while the proposed method requires centralized implementation and may be prone to single-point-of-failure risks, the one shot characteristic significantly mitigates this limitation. Importantly, unlike the existing centralized methods, such as \cite{turpin2014capt}, where agents need to communicate with the server at all times or at discrete times, the proposed approach requires computation for assigning goal positions only once. This simple fact isolates the goal assignment part from the execution phase, where the agents can track the desired trajectory independently, which makes the proposed method well-suited for precomputing and keeping the assignment solution ready for use. Additionally, through studies in Section V, it is shown that the proposed method requires very low computation in assigning goal positions even for a large swarm.

\section{Simulation Results}\label{sec:5}

In this section, the proposed goal assignment policy in Algorithm $\ref{alg:goal_assign}$ is demonstrated using MATLAB simulations through two examples, followed by a statistical analysis of the path length obtained for all point-sized agents. Additionally, Monte-Carlo studies are carried out to demonstrate the effectiveness of the proposed policy for a large swarm of disc-shaped agents. Further, the studies are extended under various practical constraints, including agent dynamics, initial position uncertainty and communication delay. The prescribed speed of the agents $v =0.5$ m/s. The parameter $\delta$ in (\ref{eq:mod_goal1}) is $0.2$. Further, we define a collision avoidance parameter $\mathcal{E}(t)$ as minimum of the distance between any two agents $i$ and $j$ at time $t$, that is, 
\begin{align}
    &\mathcal{E}(t)=\min_{\substack{i,j \in \{1, \ldots, n\} \\ i \neq j}} \mathbf{d}_{ij},& \mathbf{d}_{ij}(t)=\mid\mid\mathbf{x}_i(t) -\mathbf{x}_j(t) \mid \mid.
\end{align}
To quantify the efficiency of the resulting agents' trajectories obtained using the proposed method with respect to the shortest path to the boundary, a performance metric $\mathcal{M}_i$ is defined for the $i$th agent as the ratio of the distance covered to the shortest distance to reach $\mathcal{C}_b$, that is, 
\begin{align}\label{eq:path_lengthM}
    \mathcal{M}_i=\dfrac{\mathbf{g}_i-\mathbf{x}_{i0}}{R-\mid \mid \mathbf{x}_{i0} -\mathbf{x}_c\mid \mid}.
\end{align}
Here, $\mathcal{M}_i=1 \implies$ $\overline{\mathbf{x}_{i0}\mathbf{g}_i}$ is along the radial line $\overline{\mathbf{x}_c\mathbf{x}_{i0}}$. A higher value of $\mathcal{M}_i$ indicates a greater deviation from the shortest path to $\mathcal{C}_b$. To assess path efficiency for the swarm, another parameter $\mathcal{S}_m$ is defined as the relative increase in the sum of the path lengths for all agents compared to the sum of the shortest possible path lengths for all agents, that is,
\begin{align}\label{eq:sm}
\mathcal{S}_m=\left(\dfrac{\sum_{i=1}^{n}\mid \mid\mathbf{g}_i-\mathbf{x}_{i0}\mid \mid}{\sum_{i=1}^{n}\left(R-\mid \mid \mathbf{x}_{i0} -\mathbf{x}_c\mid \mid \right)} - 1\right ).
\end{align}

\subsection{Example 1}

\begin{figure}[!hbt]
    \centering
    \includegraphics[width=\columnwidth]{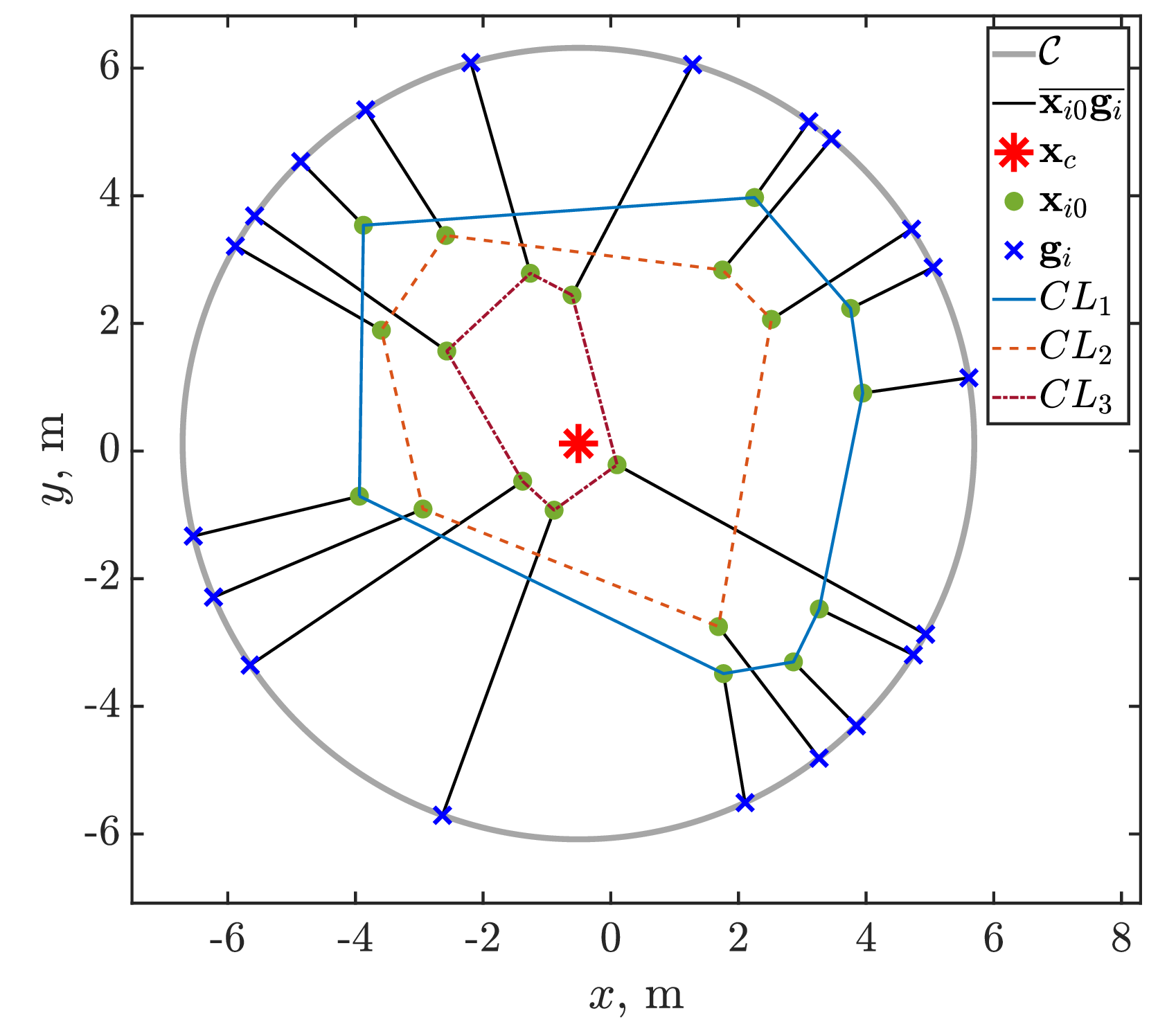}
    \caption{Example 1: Goal positions and resulting paths for 20 agents.}
    \label{fig:sim_conv_layer}
\end{figure}

\begin{figure}[!hbt]
    \centering
    \includegraphics[width=\columnwidth]{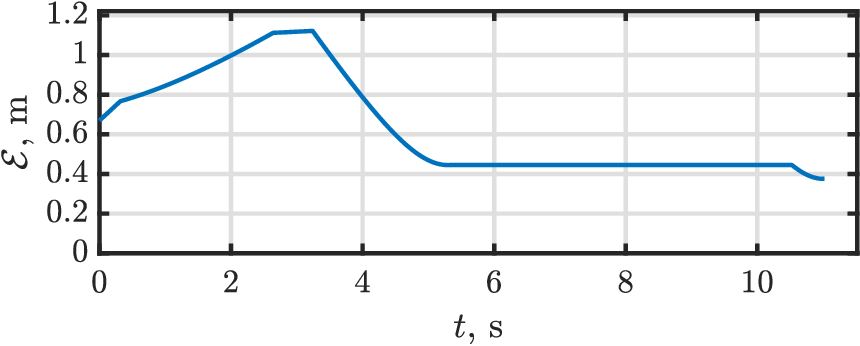}
    \caption{Variation of $\mathcal{E}$ with time for Example 1.}
    \label{fig:idc}
\end{figure}

\begin{figure}[!hbt]
    \centering
    \includegraphics[width=\columnwidth]{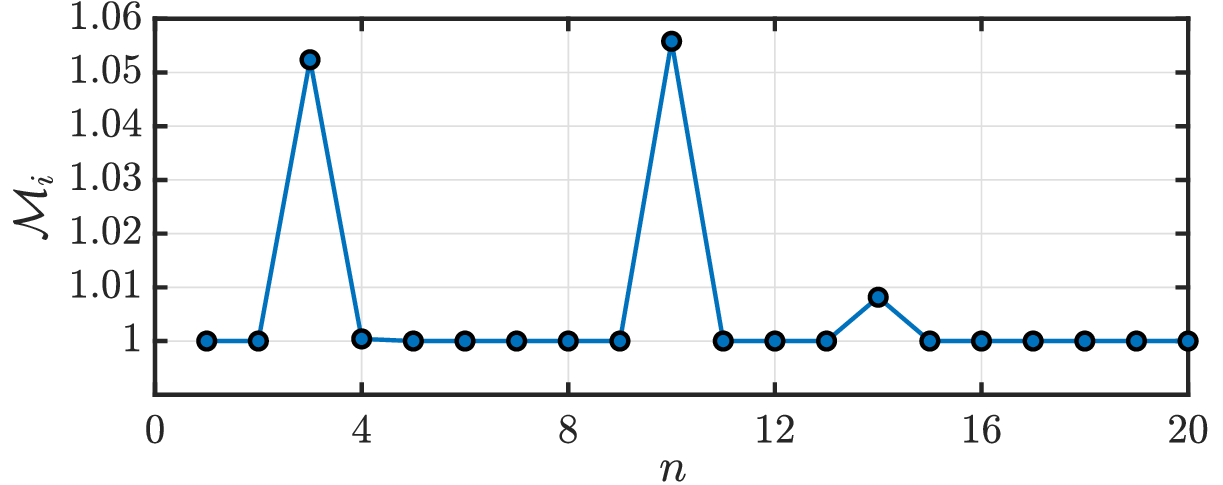}
    \caption{Variation of $\mathcal{M}_i$ for different agents in Example 1.}
    \label{fig:M_case1}
\end{figure}

This example scenario considers 20 agents with initial positions chosen randomly within a rectangular region satisfying $ x,y\in [-4,4]$. The centre of $\mathcal{C}$ is (0.15, 0.06) and its radius is 5.03 m. The set of convex layers $CL =\{CL_1,CL_2,CL_3\}$ and the unique goal position for each agent are shown in Fig. \ref{fig:sim_conv_layer}. The time at which all agents reach $\mathcal{C}_b$, that is, $\max(t_f^1,t_f^2,\dotsc,t_f^{20})=11.02$ s. The $i$th agent $(1\leq i \leq 20)$ moves towards its respective goal position along the straight path connecting its initial and goal positions as shown in Fig. \ref{fig:sim_conv_layer} with its prescribed speed, reaches its goal position $\mathbf{g}_i$ at $t = t_f^i$ and remains stationary thereafter. The time evolution of $\mathcal{E}$ for all agents is shown in Fig. \ref{fig:idc}. Here, $\mathcal{E}(t)>0$ implies that the agents do not collide with each other. For each agent, $\mathcal{M}_i$ is computed using \eqref{eq:path_lengthM} and depicted in Fig. \ref{fig:M_case1}. The average and standard deviation in $\mathcal{M}_i$ are found to be 1.006 and 0.0166, respectively. Using \eqref{eq:sm}, $\mathcal{S}_m$ is computed as 0.57\%. This implies that overall, there is an increase of 0.57\% in the sum of the path lengths of all agents relative to the sum of their optimal distances, which are along their corresponding radial direction, to the circular boundary.


\subsection{Example 2}
\begin{figure}[!hbt]
    \centering
    \includegraphics[width=\columnwidth]{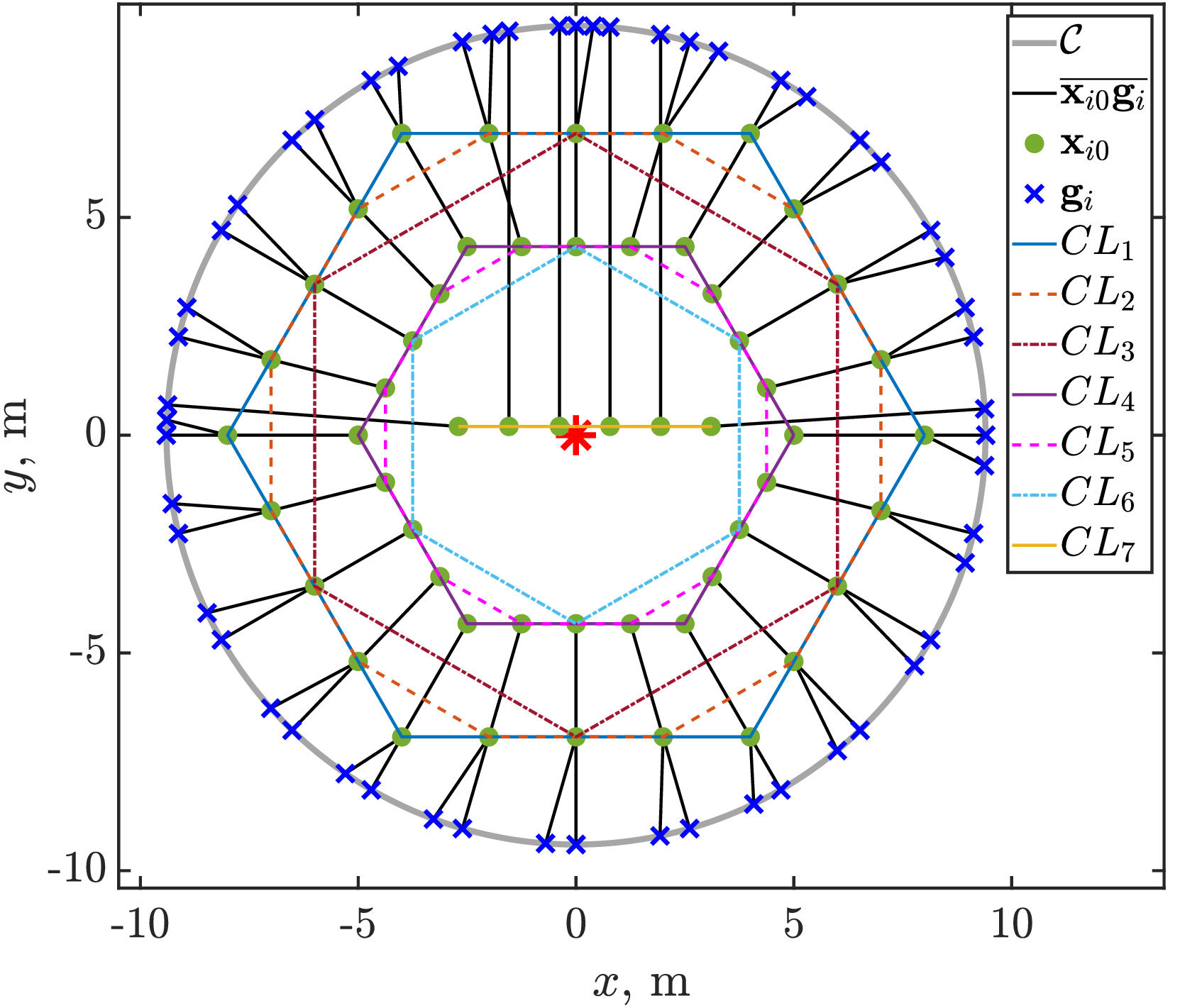}
    \caption{Example 2: Goal assignment for 54 agents.}
    \label{fig:case2_goals}
\end{figure}

\begin{figure}[!hbt]
    \centering
    \includegraphics[width=1\columnwidth]{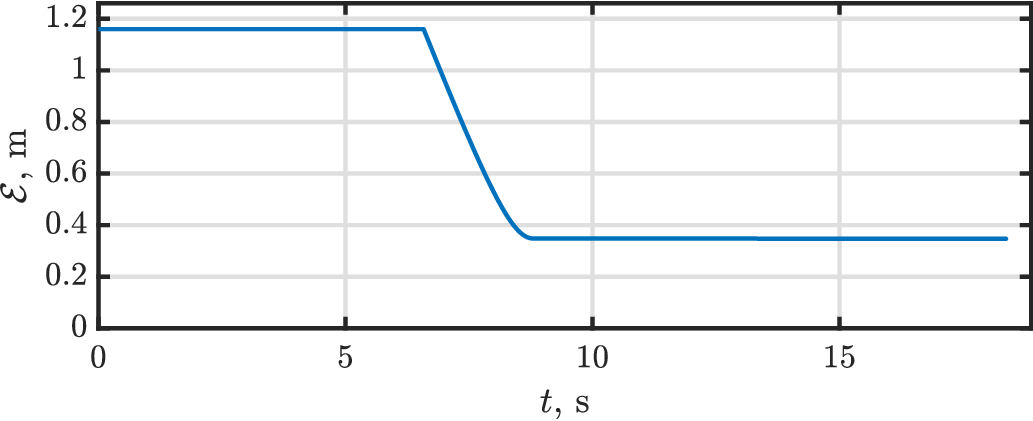}
    \caption{Variation of $\mathcal{E}$ with time for Example 2.}
    \label{fig:idc2}
\end{figure}

\begin{figure}[!hbt]
    \centering
    \includegraphics[width=\columnwidth]{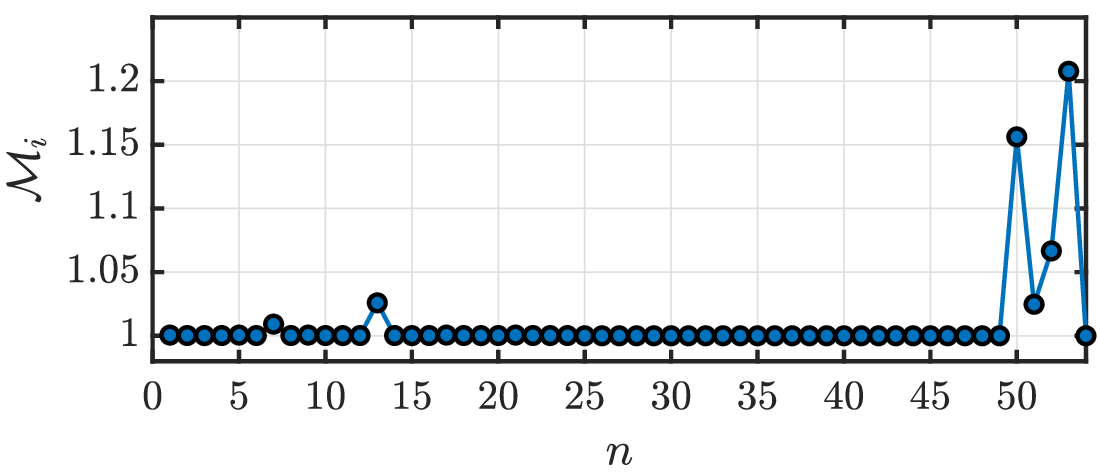}
    \caption{Variation of $\mathcal{M}_i$ for different agents in Example 2.}
    \label{fig:M_case2}
\end{figure}

   In this example, an initial arrangement of 54 agents distributed evenly on two nested regular hexagons, with side lengths 8 and 6 m, and a line segment joining [-2.9,0] and [2.9,0] is considered. Therein, 24 agents are equispaced along the perimeter of each hexagon, and the remaining 6 agents are equispaced on the line segment. The centre and radius of $\mathcal{C}$ are [0,0] and $9.4$ m, respectively. A set of seven convex layers, that is, $CL = \{CL_1,CL_2,\ldots,CL_7\}$ is formed using the initial positions of the agents. Fig. \ref{fig:case2_goals} shows the path followed by each agent to its respective goal position. The time taken for all agents to reach $\mathcal{C}_b$ is 18.4 s. The variation of $\mathcal{E}$ with time for all agents in Fig. \ref{fig:idc2} shows no inter-agent collision. For all agents, Fig. \ref{fig:M_case2} shows the values of $\mathcal{M}_i$. Note that the agents on $CL_1,~CL_2$ and $CL_3$, while deviating from the radial direction to avoid conflicting goal positions, have insignificant deviation from their shortest distance to the circular boundary. The peaks in Fig. \ref{fig:M_case2} are primarily due to the collinear agents on the innermost layer, as the search space assigned to these agents lies along the normal vector to the line connecting them, as shown in Fig. \ref{fig:search space}c. Further, the average and standard deviation of $\mathcal{M}_i$ for all agents are 1.009 and 0.036, respectively. For this example, using \eqref{eq:sm}, $\mathcal{S}_m = 1.72\%$.

 \begin{table*}[!hbt]
  \centering
  \caption{Comparison of computational time between the LCM-based approach and the proposed approach}
  \label{table:summary2}
  \begin{tabular}{c|c|c|c|c|c|c|c}
    \toprule
    $(n,R_c) $ & (10, 5m)& (50, 10m)& (100, 20m)& (500, 35m)& (1000, 50m)& (5000, 80m)& (10000, 100m)\\
    \midrule
    LCM cycle approach \cite{flocchini2017distributed} &0.00021 s& 0.0046 s & 0.0322 s & 1.3518 s& 6.923 s& 224.05 s& 2340.2 s \\
    Proposed policy & 0.0015 s& 0.0026 s & 0.004 s& 0.013 s& 0.0285 s &  0.2464 s& 0.7125 s\\
    \bottomrule
  \end{tabular}
\end{table*}

\subsection{Monte Carlo Simulations}

To quantitatively investigate the efficiency of the proposed goal assignment policy, Monte Carlo method is used. The configuration of the processor used for simulation is Intel(R) i7-9700 CPU with the clock speed of 3.00 GHz and 8 Logical Processors. Here, for different numbers of agents $n$, their initial positions are randomly sampled for 1000 test cases within a circle of radius $R_c$ centred at (0,0). The computation time of the proposed policy and LCM cycle-based approach \cite{flocchini2017distributed} is found for each test case, and their averages across all 100 test cases are listed in Table \ref{table:summary2} for different scenarios $(n,R_c)$. In contrast to the LCM cycle approach, as the number of agents and $R_c$ increase, the average computation time $T_c$ remains significantly low for the proposed policy. This can be attributed to its one-shot, conflict-free assignment characteristic, which, unlike the LCM approach, does not require computation at regular intervals.

\begin{figure}[!hbt]
    \centering
    \includegraphics[width=\columnwidth]{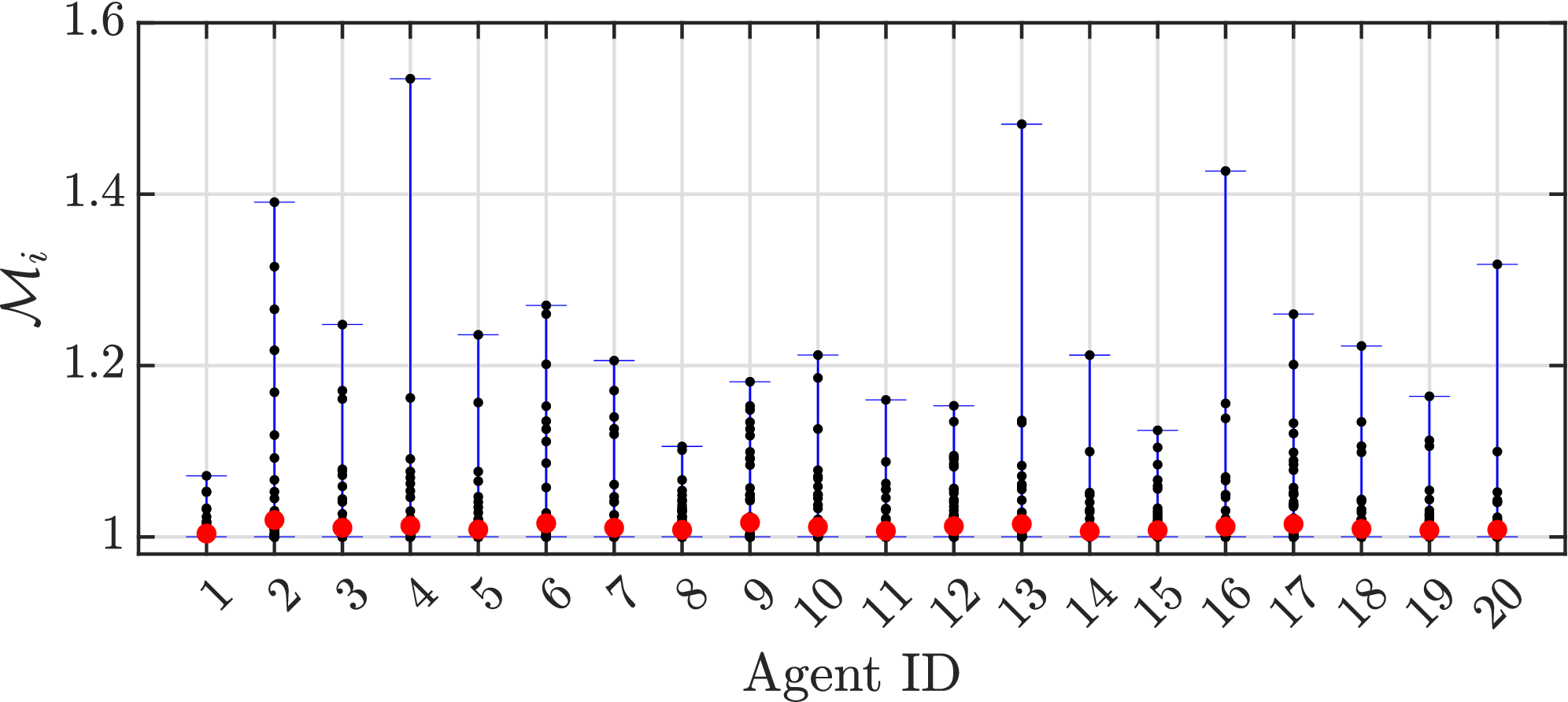}
    \caption{Monte Carlo simulation: Variation of $\mathcal{M}_i$ for 100 initial agent configurations.}
    \label{fig:rpr}
\end{figure}

Another study is carried out, analyzing path length using $\mathcal{M}_i$ of the agents. Therein, the initial positions of 20 agents are randomly sampled within the region considered in Example 1 for 100 test cases. Fig. \ref{fig:rpr} shows the values of $\mathcal{M}_i,~ \forall i =1,2,\ldots,20$, represented by black dots, while the red dot indicates the mean value of $\mathcal{M}_i$. It can be observed that the mean of $\mathcal{M}_i$ remains close to 1 for all agents. The results are summarized in Table \ref{table:summary}. Out of 100 scenarios, $\mathcal{S}_m$ is found to be less than 5 \% in 97 of them, indicating that in the majority of the test cases, the paths of the agents are along the shortest path to the circular boundary.
 \begin{table}[!hbt]
  \centering
  \caption{Summary of path length analysis}
  \label{table:summary}
  \begin{tabular}{cc|cc}
    \toprule
    $\mathcal{S}_m $ & \textbf{\#test cases} & $\mathcal{S}_m $ & \textbf{\#test cases} \\
    \midrule
     $0<\mathcal{S}_m\leq0.01$ &45 &$0.03<\mathcal{S}_m\leq 0.04$ & 7  \\
    $0.01<\mathcal{S}_m\leq0.02$ & 31 &$0.04<\mathcal{S}_m\leq0.05$ & 4 \\
    $0.02<\mathcal{S}_m\leq0.03$ & 10 &$0.05<\mathcal{S}_m$ & 3\\
    \bottomrule
  \end{tabular}
\end{table}

\subsection{Monte Carlo Study Considering Disc-Shaped Agents}\label{sec:6d}

Results presented thus far in the paper involve point-sized agents. In this study, the agents are considered to be disc-shaped with their dimensions being modeled on Crazyflie 2.0 quadrotor \cite{giernacki2017crazyflie}. Accordingly, each agent can fit within a disc of diameter 15 cm and the collision avoidance condition in \eqref{eq:problem_goal2} is modified as:
\begin{equation}\label{eq:conf_cond}
    \mid \mid \mathbf{x}_i(t)- \mathbf{x}_j(t)\mid \mid>d_s,~ \forall i \neq j,0<t\leq\max(t_f^i,t_f^j),
\end{equation}
where $d_s=0.15$ m is the safety distance among the agents. A conflict between agents $i$ and $j$ is said to occur if the condition in \eqref{eq:conf_cond} is violated. Further, the constant $\delta$ in \eqref{eq:mod_goal1} is considered as 0.5. A Monte Carlo study is carried out, where the positions of $n$ agents are randomly sampled for 1000 test cases within a circle $\mathcal{C}_{\text{cir}}$ centered at $(0,0)$ and radius $R_c$, while ensuring that the initial separation between any two agents is at least 0.4 m. For a given scenario $(n,R_c)$, let $\mathcal{I}_{col}$ be the set of test case(s) which have conflict(s). To evaluate the effectiveness in finding a conflict-free path for all agents, we define $\mathcal{P}_{col}$ as
\begin{equation}\label{eq:pc_prob}
    \mathcal{P}_{col}(n,R_c) = \dfrac{|\mathcal{I}_{col}|}{\# \text{Test cases}},
\end{equation}
where $|\mathcal{I}_{col}|$ is the cardinality of the set $\mathcal{I}_{col}$. Using \eqref{eq:pc_prob}, $\mathcal{P}_{col} = 0$ corresponds to no test cases having conflict(s). Let $N_{col}^i$ be the number of conflict(s) in the $i$th test case of a given scenario. Further, we define $\mu_{col}$, $\sigma_{col}$ and $N^{\max}_{col}$ as the average number, standard deviation and maximum number, respectively, of conflict(s) observed in $\mathcal{I}_{col}$, that is,
\begin{align}
    \mu_{col} &= \begin{cases}
        \dfrac{\sum_{i \in \mathcal{I}_{col}} N_{col}^i }{|\mathcal{I}_{col}|},& \text{if } |\mathcal{I}_{col}|\neq0\\
        0,& \text{otherwise}
    \end{cases}\\
    \sigma_{col}& =\begin{cases}
        \sqrt{\dfrac{\sum_{i \in \mathcal{I}_{col}}\left(N_{col}^i-\mu_{col}\right )^2} {|\mathcal{I}_{col}|}},& \text{if } |\mathcal{I}_{col}|\neq0\\
        0,& \text{otherwise}
    \end{cases}\\
    N^{\max}_{col} &= \max_{i \in \mathcal{I}_{col}}N_{col}^i.
\end{align}
To analyze the path lengths across the test cases, a parameter $\mathcal{S}_m^{avg}$ is defined as the mean of $\mathcal{S}_m$ over all test cases, that is,
\begin{equation}
  \mathcal{S}_m^{avg} = \frac{1}{N} \sum_{i=1}^{N} \mathcal{S}_{m,i},
\quad i \in {1,2,\ldots,N},
\end{equation} where $\mathcal{S}_{m,i}$ is the value for $\mathcal{S}_m$ for the $i$th test case and $N$ is the total number of test cases.

\begin{table*}[!hbt]
  \centering
  \caption{Conflict analysis considering disc-shaped agents for specific pairs of $(n,R_c)$.}
  \label{table:col_sc}
\setlength{\tabcolsep}{2.0pt} 
\renewcommand{\arraystretch}{1.2} 
\begin{tabularx}{\textwidth}{@{}c|*{3}{>{\centering\arraybackslash}X}|*{3}{>{\centering\arraybackslash}X}|*{3}{>{\centering\arraybackslash}X}@{}}
\toprule
\multirow{2}{*}{$n$} & \multicolumn{3}{c|}{$ R_c =40 $ m} & \multicolumn{3}{c|}{$R_c =50$ m} & \multicolumn{3}{c}{$R_c =60$ m} \\ 
\cmidrule(lr){2-4} \cmidrule(lr){5-7} \cmidrule(lr){8-10} 
& $\mathcal{P}_{col}$ & $(\mu_{col},\sigma_{col},N^{\max}_{col})$ &$~~\mathcal{S}_m^{avg}$, \% & $\mathcal{P}_{col}$ & $(\mu_{col},\sigma_{col} ,N^{\max}_{col})$ &$~~\mathcal{S}_m^{avg}$, \%& $\mathcal{P}_{col}$ & $(\mu_{col} ,\sigma_{col},N^{\max}_{col})$&$~~\mathcal{S}_m^{avg}$, \% \\ 
\midrule
10 & 0 & (0, 0, 0) & 3.60 & 0 & (0, 0, 0) & 3.78 & 0 & (0, 0, 0) & 4.08 \\
20 & 0 & (0, 0, 0) & 1.92 & 0 & (0, 0, 0) & 2.06 & 0 & (0, 0, 0) & 2.12 \\
30 & 0 & (0, 0, 0) & 1.16 & 0 & (0, 0, 0) & 1.38 & 0 & (0, 0, 0) & 1.41 \\
40 & 0.004 & (1, 0, 1) & 0.83 & 0.002 & (1, 0, 1) & 0.91 & 0 & (0, 0, 0) & 0.99 \\
50 & 0.008 & (1, 0, 1) & 0.58 & 0.003 & (1, 0, 1) & 0.67 & 0.001 & (1, 0, 1) & 0.74 \\
60 & 0.016 & (1, 0, 1) & 0.44 & 0.005 & (1, 0, 1) & 0.52 & 0.003 & (1, 0, 0) & 0.55 \\
70 & 0.018 & (1, 0, 1) & 0.36 & 0.014 & (1, 0, 1) & 0.44 & 0.007 & (1, 0, 1) & 0.45 \\
80 & 0.028 & (1, 0, 1) & 0.27 & 0.017 & (1, 0, 1) & 0.34 & 0.012 & (1, 0, 1) & 0.39 \\
90 & 0.033 & (1, 0, 1) & 0.25 & 0.021 & (1, 0, 1) & 0.28 & 0.013 & (1, 0, 1) & 0.32 \\
100 & 0.036 & (1, 0, 1) & 0.21 & 0.023 & (1, 0, 1) & 0.25 & 0.015 & (1, 0, 1) & 0.28 \\
\bottomrule
\end{tabularx}
\end{table*}

\begin{table*}[!hbt]
  \centering
  \caption{Conflict analysis and average computation time considering a large swarm of disc-shaped agents for different $(n,R_c)$.}
  \label{table:summary3}
  \begin{tabular}{c|c|c|c|c|c|c}
    \toprule
    $(n,R_c) $ & (500, 150 m)& (500, 200 m)& (1000, 150 m)& (1000, 200 m)& (2000, 150 m)& (2000, 200 m)\\
    \midrule
    $\mathcal{P}_{col}$ & 0.005 & 0.002  & 0.031 & 0.018  &  0.044 & 0.021 \\
    $(\mu_{col},\sigma_{col},N^{\max}_{col})$&  (1,0,1) & (1,0,1) &(1.055,0.23,2) & (1.032,0.18,2) &  (1.068, 0.25,2)&(1.037,0.19,2) \\
    $\mathcal{S}_m^{avg}$, \% & 0.176& 0.208 & 0.075 & 0.096 & 0.025 & 0.0036 \\
    \bottomrule
  \end{tabular}
\end{table*}

Table \ref{table:col_sc} presents the results of the studies carried out for many scenarios, that is, for specific pairs of $(n,R_c)$. Therein, consider an example scenario with $n=100,$ $R_c=40$ m. As shown in Table \ref{table:col_sc}, with $\mathcal{P}_{col}=0.036$ using \eqref{eq:pc_prob}, the number of test cases having conflict(s) = $\mathcal{P}_{col}\times 1000=36$. Further, $\mu_{col}=1,\sigma_{col}=0$ indicates that the number of conflicts in each of these 36 test cases is exactly 1. Based on Table \ref{table:col_sc}, following observations can be drawn:
\begin{enumerate}
    \item For small values of $n (\leq 30)$, the number of test cases with conflicts is 0.
    \item For moderate values of $n(40\leq n \leq 70)$, the number of test cases with conflicts is at most 18. Therein, $\mu_{col} =1,\sigma_{col}=0$ and $ N^{\max}_{col}=1$ imply that there is only 1 conflict in any of the test cases that have conflict(s).
    \item For high values of $n(80\leq n \leq 100)$, at most 36 out of the 1000 test cases have conflict(s). Again, $\mu_{col} =1,\sigma_{col}=0$ and $N^{\max}_{col}=1$ indicate that the number of conflict(s) in each of those cases is 1.
    \item For all scenarios $(n,R_c)$, $\mathcal{S}_m^{avg}$, defined as the mean of $\mathcal{S}_m$ computed across the 1000 test cases, is less than 4.08\%. 
\end{enumerate}

The studies are further extended to a much higher number of disc-shaped agents. Table \ref{table:summary3} presents the results of the Monte-Carlo study comprising 1000 test cases each for the scenarios $(n,R_c) =\{(500, 150 \text{ m}), (500,200 \text{ m}),(1000,150\text{ m}),(1000,200\text{ m}),$ $(2000,150\text{ m}), (2000,200\text{ m})\}$ and it also includes $\mathcal{S}_m^{avg}$, which computes the average of $\mathcal{S}_m$ over the 1000 test cases, for each scenario. It can be observed that for all $(n,R_c)$ considered in Table \ref{table:summary3}, $\mathcal{P}_{col}\leq 0.057$ and $\mathcal{S}_m^{avg}\leq 0.208$ \%.

While the proposed goal assignment policy does not guarantee inter-agent collision avoidance for disc-shaped agents, studies in this subsection show that it effectively finds conflict-free paths for all agents in nearly all test cases across the scenarios. Even among the test cases that have conflicts, it is found from Tables \ref{table:col_sc} and \ref{table:summary3} that the number of conflicts is limited to 1 for moderate number of agents $(n\leq 100)$ and 2 for high number of agents $(n\geq 500)$. The corresponding conflicts encountered by the agents can be easily detected, for example, using the point of closest approach method \cite{dunthorne2014estimation}. Further, in a distributed manner, such conflicts can be resolved by incorporating speed assignment policies, such as in \cite{singh2018reactive,alejo2012speed}. The path length analysis as characterized by $\mathcal{S}_m^{avg}$ in Tables \ref{table:col_sc} and \ref{table:summary3} also demonstrates that the proposed method assigns the goal positions to agents such that the distance between their initial and goal positions is close to the shortest distance from their initial positions to the circular boundary.

\subsection{Performance Under Real-World Considerations}

This study presents a robustness analysis of the proposed method in finding conflict-free paths for disc-shaped agents while considering realistic constraints, such as agent dynamics, uncertainty in initial position measurement and communication delays. Subsequently, we define the practical attributes incorporated in the study.

\subsubsection{Agent Dynamics} 
\begin{figure}[!hbt]
    \centering
    \includegraphics[width=0.8\columnwidth]{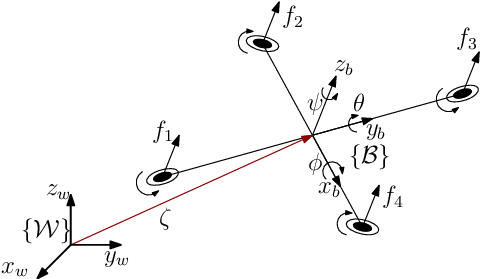}
    \caption{Quadrotor model schematic.}
    \label{fig:quad_model}
\end{figure}

The agent dynamics is modeled using a linearized model of an underactuated six degrees of freedom (dof) quadrotor \cite{beard2008quadrotor}. The coordinate system in the inertial $\{\mathcal{O}\}$ and body $\{\mathcal{B}\}$ frame, and the free body diagram of the quadrotor are shown in Fig. \ref{fig:quad_model}. 
The position vector of the quadrotor relative to $\{\mathcal{O}\}$ is denoted by $\zeta(t) = [x(t),y(t),z(t)]^T$. The attitude vector consisting of roll $\phi$, pitch $\theta$ and yaw $\psi$ is represented by $\Omega = [\phi,\theta,\psi]^T$. The Euler angle rotation matrix from $\{\mathcal{B}\}$ to $\{\mathcal{O}\}$ using $\mathrm{ZYX}$ convention is governed by
\begin{equation}\label{eq:rot_matrix}
    \mathcal{R}_{bw} = \begin{bmatrix}        
c \theta c \psi & s \phi s \theta c \psi-c \phi s \psi & c \phi s \theta c \psi+s \phi s \psi \\
c \theta s \psi & s \phi s \theta s \psi+c \phi c \psi & c \phi s \theta s \psi-s \phi c \psi \\
-s \theta & s \phi c \theta & c \phi c \theta
    \end{bmatrix},
\end{equation}
where $c \phi \triangleq \cos \phi,c \theta \triangleq \cos \theta,c \psi \triangleq \cos \psi \text { and } s \phi \triangleq \sin \phi,s \theta \triangleq \sin \theta,s \psi \triangleq \sin \psi$. Let $(u,v,w)$ and $(p,q,r)$ denote the linear velocities and angular velocities in the body frame. The relationship between the position vector and the body frame velocities is given by
\begin{align}
    \dot{\zeta} =\begin{bmatrix}
        \dot{x}(t)\\
        \dot{y}(t)\\
        \dot{z}(t)
    \end{bmatrix} = \mathcal{R}_{bw}\begin{bmatrix}
        u\\
       v\\
        w
    \end{bmatrix}.
\end{align}
The relation between attitude angle rates $(\dot{\phi},\dot{\theta},\dot{\psi})$ and body frame velocities $(p,q,r)$ is governed by
\begin{align}
   \begin{bmatrix}
       \dot{\phi}\\ \dot{\theta}\\ \dot{\psi}
   \end{bmatrix} =\begin{bmatrix}
1 & \sin (\phi) \tan (\theta) & \cos (\phi) \tan (\theta) \\
0 & \cos (\phi) & -\sin (\phi) \\
0 & \sin (\phi) \sec (\theta) & \cos (\phi) \sec (\theta)
\end{bmatrix}\begin{bmatrix}
    p\\q\\r
\end{bmatrix}.
\end{align}
Assuming that $\phi,\theta $ are small, the Coriolis terms in the quadrotor dynamics can be neglected, $\dot{\Omega} \approx [p,q,r]^T$, and a simplified inertial model is obtained. Accordingly, the translational and rotational dynamics of the quadrotor are given by 
\begin{align}
    &\begin{bmatrix}
       \ddot{x}\\
       \ddot{y}\\
       \ddot{z}
    \end{bmatrix} =\begin{bmatrix}
        0\\0\\-g
    \end{bmatrix} + \mathcal{R}_{bw} \begin{bmatrix}
        0\\0\\F
    \end{bmatrix},\\
  &  \begin{bmatrix}
       \ddot{\phi}\\
       \ddot{\theta}\\
       \ddot{\psi}
    \end{bmatrix} =\begin{bmatrix}
        \tau_{\phi}/J_x\\ 
        \tau_{\theta}/J_y \\
        \tau_{\psi}/J_z
    \end{bmatrix},
\end{align}
where $F= (f_1+f_2+f_3+f_4)$ is the sum of thrust generated by the four rotors in the body $z$-direction, $( \tau_{\phi},\tau_{\theta},\tau_{\psi})$ are the rolling, pitching and yawing torque, respectively, in the body frame, and $(J_x,J_y,J_z)$ are the principal moments of inertia of the quadrotor about the body $x$-, $y$- and $z$-axes, respectively. Further, a cascaded proportional-derivative control architecture is utilized to deduce the control inputs $(F,\tau_{\phi},\tau_{\theta},\tau_{\psi})$, where the outer loop controller tracks the desired trajectory by generating desired pitch and roll angles, and the inner loop controller controls the attitude angle accordingly \cite{subudhi2018modeling}. The outer loop controller governs the commanded acceleration, and the associated equations are as follows:
\begin{align}\label{eq:outer_loop}
    \begin{bmatrix}
        \ddot{x}_c\\
        \ddot{y}_c\\
        \ddot{z}_c
    \end{bmatrix}
    =
    \mathbf{K_{D}}
    \begin{bmatrix}
        \dot{x}_d - \dot{x}\\
        \dot{y}_d - \dot{y}\\
        \dot{z}_d - \dot{z}
    \end{bmatrix}
    +
    \mathbf{K_{P}}
    \begin{bmatrix}
        x_d - x\\
        y_d - y\\
        z_d - z
    \end{bmatrix},
\end{align}
where $\mathbf{K}_P= \mathrm{diag}(k_{p_x}, k_{p_y}, k_{p_z})$ and $\mathbf{K}_D=  \mathrm{diag}(k_{d_x}, k_{d_y}, k_{d_z}) $ are the proportional and derivative gain matrices for the outer loop, respectively. The desired position $(x_d,y_d,z_d)$ and velocity $(\dot{x}_d, \dot{y}_d, \dot{z}_d)$ are required for commanding the quadrotor. In this work, $z_d = 1$ m and $\dot{z}_d = 0$ for all agents. Further, for the $i$th agent, the desired velocity components $\dot{x}_d=v\cos\psi_i$ and $\dot{y}_d = v\sin \psi_i$ are obtained using \eqref{eq:current_rel1}, based on which the desired position components $x_d$ and $y_d$ are computed as follows:
\begin{align}
    \begin{bmatrix}
        x_d(t),&
        y_d(t)
    \end{bmatrix} = \mathbf{x}_{i0} +vt\begin{bmatrix}
        \cos\psi_i,&
        \sin\psi_i
    \end{bmatrix}.
\end{align}
The commanded roll and pitch angles obtained through the linearized dynamics using \cite{beard2008quadrotor} are given by
\begin{align}
    \begin{bmatrix}
        \phi_c\\ \theta_c
    \end{bmatrix} = \begin{bmatrix}
        \sin \psi_{d} & \cos \psi_{d}\\
        -\cos \psi_{d} & \sin \psi_{d}
    \end{bmatrix} \begin{bmatrix}
        \ddot{x}_c \\ \ddot{y}_c
    \end{bmatrix},
\end{align}
where $\psi_c =\psi_d $ is the commanded/desired yaw angle and the quadrotor is assumed to have a fixed heading during its motion, that is, $\psi_c =\psi_d=0$. The inner loop relates the attitude angle control to the rolling, pitching and yawing torque, and the total thrust is governed by the commanded acceleration in $z-$direction as follows:
\begin{align}
    F &= m(g +\ddot{z}_c),\\
   \begin{bmatrix}
        \tau_{\phi}\\
        \tau_{\theta} \\
        \tau_{\psi} 
    \end{bmatrix}  &=\mathbf{K_{D\tau}}
    \begin{bmatrix}
        p_c - p\\
        q_c - q\\
        r_c - r
    \end{bmatrix}
    +
    \mathbf{K_{P\tau}}
    \begin{bmatrix}
        \phi_c- \phi\\
       \theta_c - \theta\\
        \psi_c - \psi
    \end{bmatrix},\label{eq:rot_dyn}
\end{align}
where $m$ is the mass of the quadrotor, $g =9.81$ m/s$^2$ is the acceleration due to gravity, and $\mathbf{K}_{P\tau}= \mathrm{diag}(k_{p\tau_{\phi}}, k_{p\tau_{\theta}}, k_{p\tau_{\psi}})$ and $\mathbf{K}_{D\tau}=  \mathrm{diag}(k_{d\tau_\phi}, k_{d\tau_{\theta}}, k_{d\tau_{\psi}}) $ are the proportional and derivative gain matrices for the inner loop, respectively. Further, we set $p_c = q_c = r_c = 0$ to hold the quadrotor at rest at the desired orientation.

\subsubsection{Position Uncertainty} Since the proposed method utilizes only the initial position information of the agents, this attribute considers that the initial position of the agents, known to the central server computing goal positions for all agents, is uncertain. Accordingly, for the $i$th agent, the initial position known to the central server is considered as
\begin{align}\label{eq:pos_uncertainty}
    &\Tilde{\mathbf{x}}_i =  \mathbf{x}_i +\Delta_u, & \Delta_u \sim \mathcal{U}(-\delta_u,\delta_u)^2
\end{align}
where $\Delta_u\in \mathbb{R}^2$ is the random perturbation vector with each element of $\Delta_u$ being uniformly sampled between $(-\delta_u,\delta_u)$. Here, $\mathcal{U}(a,b)$ denotes the uniform distribution between two real numbers $a$ and $b$. The measured position $\tilde{\mathbf{x}}_i$ (in place of $\mathbf{x}_i$) is then plugged into Algorithms \ref{alg:convex_layer} and \ref{alg:goal_assign} to obtain goal positions for all agents.

\subsubsection{Communication delays} This attribute considers a delay in communicating the respective goal positions to the agents by the central server. The agents initiate their motion only at the instant when the goal position information is received. Accordingly, this attribute leads to different start times for agents as they move towards their respective goal positions. The time delay for the $i$th agent, $\Delta_t^i$ is considered as a random variable uniformly distributed between 0 and $\delta_{td}$, that is, $\Delta_t^i \sim \mathcal{U}(0,\delta_{td})$, where $\delta_{td}$ is the maximum delay. Using \eqref{eq:current_rel1}, the position of the $i$th agent considering the delay $\Delta_t^i$ is governed by
\begin{align}
    \mathbf{x}_i(t) = 
    \begin{cases}
        \mathbf{x}_{i0}, & \text{if } t \in [0, \Delta_t^i) \\
        \mathbf{x}_{i0} + v\left(t - \Delta_t^i\right)
        \begin{bmatrix}
            \cos \psi_i \\
            \sin \psi_i
        \end{bmatrix}^T, & \text{if } t \in [\Delta_t^i, t_f^i + \Delta_t^i).
    \end{cases}
\end{align}
Studies are carried out considering different combinations of realistic attributes as described in Table \ref{table:cond_def}. The gain values in \eqref{eq:outer_loop} and \eqref{eq:rot_dyn} are as follows: $\mathbf{K_P}=\mathrm{diag}(7.76,6.46,7.02)$, $\mathbf{K_D}=\mathrm{diag}(4.56,4.16,5.16)$, $\mathbf{K_P}_{\tau}=\mathrm{diag}(4.33,3.45,4.02)$, $\mathbf{K_D}_{\tau}=\mathrm{diag}(1.59,1.16,2.37)$. The nominal values of mass and moments of inertia considered for the quadrotor are given by $m = 0.964$ kg and $\{J_x = J_y=8.55\times 10^{-3},J_z = 1.47\times 10^{-2} \}$ kg$\cdotp$m$^2$, respectively. Considering heterogeneity in the swarm, the mass and moments of inertia of each agent are independently sampled from a uniform distribution within $\pm20\%$ of their nominal values, while maintaining the physical relationship between them.
\begin{table}[!hbt]
  \centering
  \caption{Scenarios for robustness analysis.}
  \label{table:cond_def}
  \begin{tabular}{c|c|c|c}
    \toprule
    & \makecell{Agent\\dynamics} & \makecell{Communication\\delay} & \makecell{Uncertain\\initial position}\\
    \midrule
    \textbf{Baseline} & $\times$ & $\times$ & $\times$ \\
    \textbf{Only Dynamics} & $\checkmark$ & $\times$ & $\times$ \\
    \textbf{Dyn + PosErr} & $\checkmark$ & $\times$ & $\checkmark$ \\
    \textbf{Dyn + ComDelay} & $\checkmark$ & $\checkmark$ & $\times$ \\
    \textbf{All Disturbances} & $\checkmark$ & $\checkmark$ & $\checkmark$ \\
    \bottomrule
  \end{tabular}
\end{table}
\begin{figure}[!hbt]
    \centering
        \includegraphics[width=0.94\linewidth]{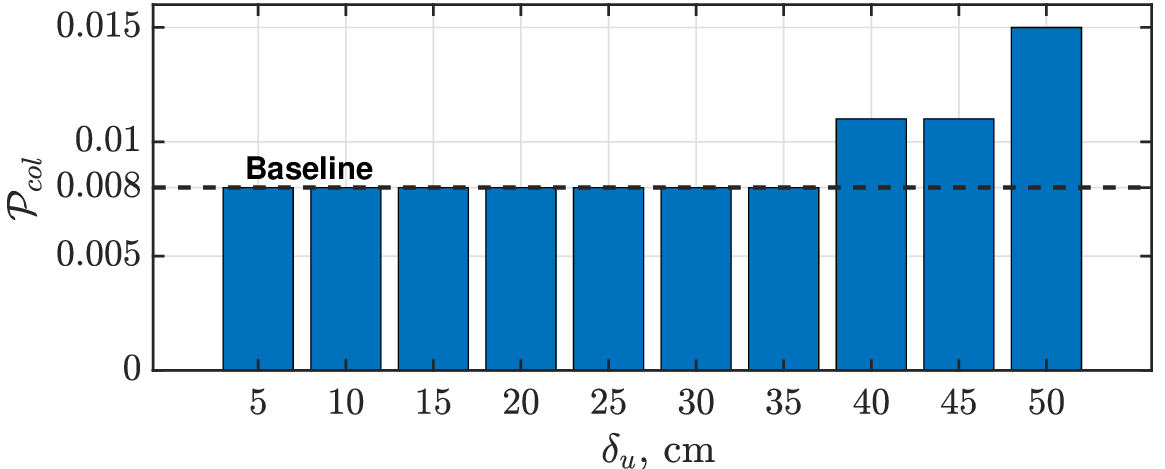} 
         \caption{Results with \textbf{Dyn + PosErr} for $(n,R_c) = (50,40\text{ m})$.}
        \label{fig:noise_data_cl}
\end{figure}

\begin{figure}[!hbt]
   \centering
        \includegraphics[width=0.94\linewidth]{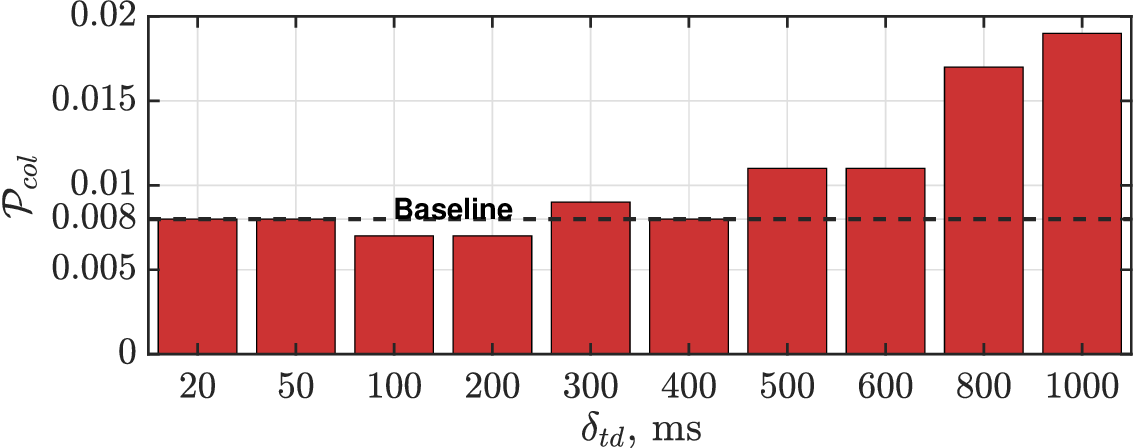} 
        \caption{Results with \textbf{Dyn + ComDelay} for $(n,R_c) = (50,40\text{ m})$.}
        \label{fig:del_data_cl}
\end{figure}

Considering the 1000 test cases from the scenario $(n,R_c) = (50,40\text{ m})$ in Table \ref{table:col_sc} and the attribute combination \textbf{Dyn + PosErr}, Fig. \ref{fig:noise_data_cl} shows the variation of $\mathcal{P}_{col}$ with respect to $\delta_u=\{5,10,\ldots,50 \}$ cm. It can be observed that, compared to \textbf{Baseline}, there is no change in $\mathcal{P}_{col}$ up to $\delta_u = 35$ cm. With the same set of 1000 test cases, Fig. \ref{fig:del_data_cl} illustrates the variation of $\mathcal{P}_{col}$ with $\delta_{td} = \{20, 50,100,200,300,400,500,600,800,1000 \}$ ms for \textbf{Dyn + ComDelay} scenario. Those results show that $\mathcal{P}_{col}$ remains lower than or equal to \textbf{Baseline} for $\delta_{td} \leq 200$ ms. Note that according to the empirical data in \cite{tardioli2017pound}, which evaluates delays in ROS-based communication in multi-agent systems, the maximum realistic delay is observed to be 200 ms, and the proposed method shows no deterioration in that range.

\begin{figure}[!hbt]
   \centering
    \includegraphics[width=0.94\columnwidth]{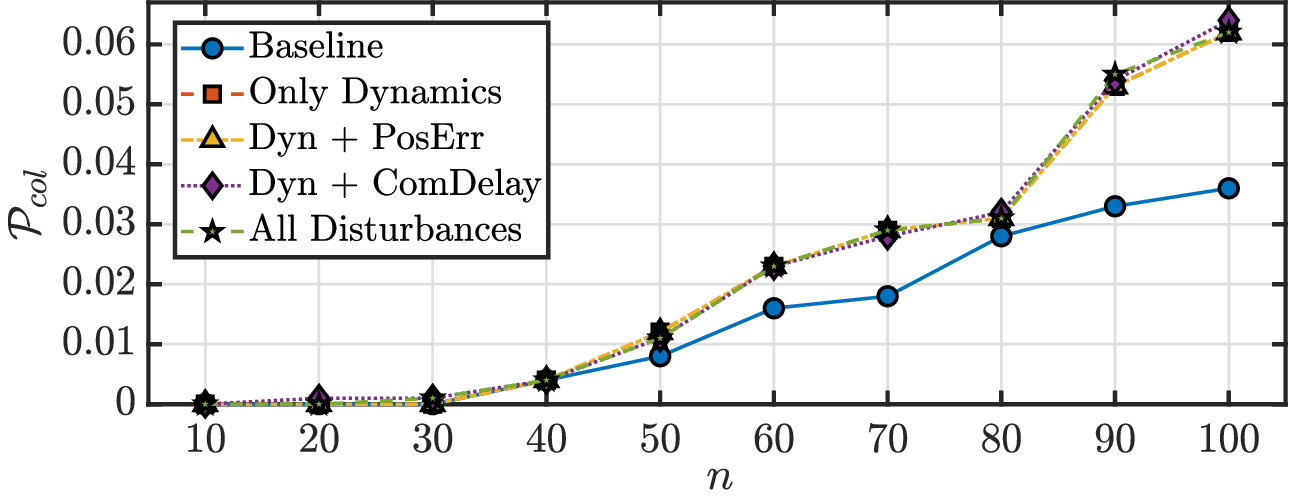}
    \caption{Robustness study with various combinations of practical attributes.}
    \label{fig:all_data}
\end{figure}

In another study with $\delta_{td} = 200$ ms and $\delta_u = 20$ cm, $\mathcal{P}_{col}$ is evaluated for different combinations of attributes in Table \ref{table:cond_def} by considering 1000 test cases with $n = \{10,20,\ldots,100 \}$ and $R_c = 40$ m from Table \ref{table:col_sc}. Fig. \ref{fig:all_data} presents the results from that study and illustrates the variation of $\mathcal{P}_{col}$ with the number of agents for different attribute combinations. While there is an increase in $\mathcal{P}_{col}$, the deterioration is not very significant and $\mathcal{P}_{col}$ across all $(n,R_c)$ scenarios remains lower than or equal to 0.064.


\section{Conclusion}\label{sec:6}

This article presents a one-shot goal assignment policy for distributing multiple point-sized agents along a circular boundary encompassing the agents. Utilizing the geometry of convex layers, a search space region is proposed for each of the agents. Regardless of the initial arrangement of agents, the proposed goal assignment policy ensures a unique goal position for each agent within its search space region. A guarantee for inter-agent collision avoidance is established using the property that a point in the search space of an agent is closer to that agent as compared to any other agent lying within or on the convex layer on which the agent lies. Further, the proposed approach facilitates the choice of any prescribed speed for agents. A qualitative comparison with the existing works also highlights the advantage in terms of computational complexity that the proposed policy offers as the problem scales. The statistical analysis in different scenarios shows that the path length of agents remains very close to the optimal distance between the agents and circular boundary. Additional Monte Carlo studies are included in this work, which consider up to 2000 randomly placed disc-sized agents of diameter 15 cm and evaluate the performance of the proposed method. Across simulations with 10 to 100 agents, the results highlight that the proposed policy effectively finds collision-free paths for at least 96.4\%, 97.3\% and 98.5\% of the 1000 test cases for circular boundaries of radii 40 m, 50 m and 60 m, respectively. For a very high number of disc-shaped agents (up to 2000), the success rate in finding collision-free paths remains more than 95.6\%. In another study, various practical attributes, such as six-dof quadrotor dynamics, uncertainty in position measurement, communication delays, and heterogeneity in quadrotor parameters, are considered for the disc-shaped agents, and a robustness analysis is performed for different combinations of these attributes. Considering all such factors, the proposed method demonstrates satisfactory performance with a success rate of at least 93.6\% across all tested scenarios.

In contrast to existing works, the proposed method offers a guaranteed solution to the problem at the initial time itself. Once the goal positions are assigned, the agents do not require any runtime replanning or communication with the central server. These advantages make the proposed method particularly relevant in the scenario of a large swarm of agents with constrained communication capabilities. Future research directions include extending the proposed strategy to agents with different dynamic models and to scenarios involving static or dynamic obstacles. Additionally, the goal assignment problem for other shapes, layers and configurations of boundaries presents another challenge in defining the search space regions and can be considered as a potential future work. Another future research direction lies in incorporating event-triggered local trajectory planning elements at the agent level, which cater to disturbances, changes in the environment or agent failures during the execution phase.

\bibliographystyle{ieeetr}
\bibliography{sample.bib}

\begin{IEEEbiography}[{\includegraphics[width=1in,height=1.25in,clip,keepaspectratio]{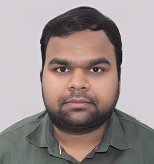}}]{Gautam Kumar}
received the B.Tech.\ degree in Mechanical Engineering from Indian Institute of Technology Indore, India, in 2020. He is currently a Ph.D.\ candidate at the Autonomous Vehicles Laboratory in the Department of Aerospace Engineering, Indian Institute of Science Bangalore, India. 

His research interests include multi-agent systems, autonomous UAV coordination, and guidance and control.
\end{IEEEbiography}

\begin{IEEEbiography}[{\includegraphics[width=1in,height=1.25in,clip,keepaspectratio]{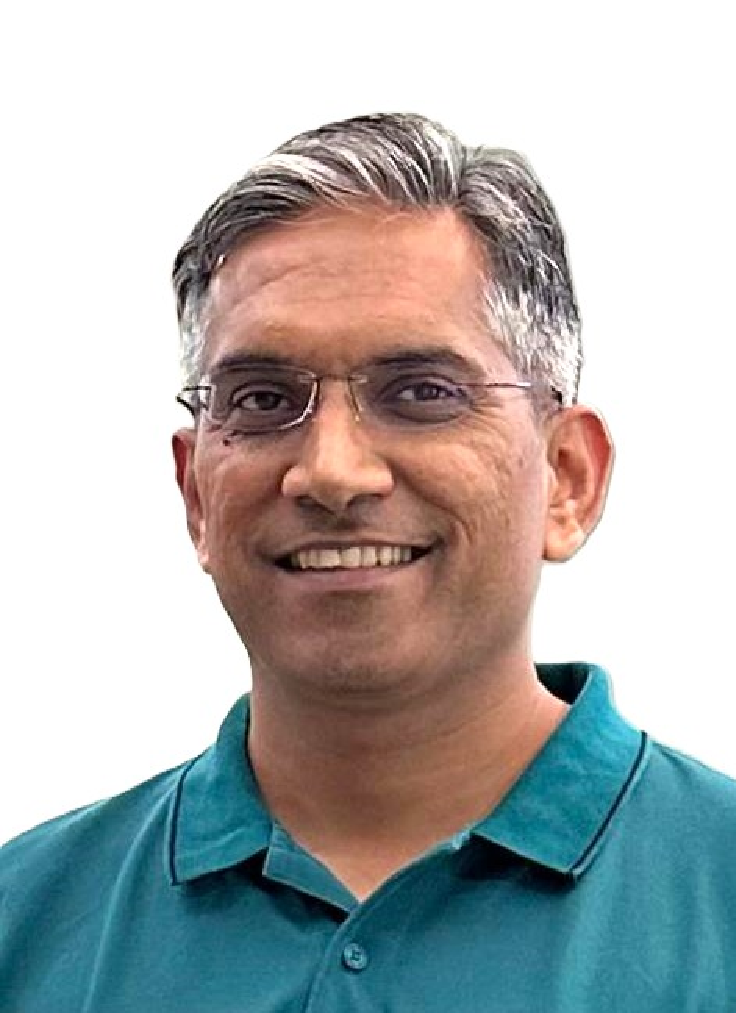}}]{Ashwini Ratnoo}

received the B.E. degree in Electrical Engineering from the MBM Engineering College, Jodhpur, India, in 2003, and the M.E. and Ph.D. degrees, in Aerospace Engineering, from the Indian Institute of Science, Bangalore, India, in 2005 and 2009, respectively. During 2009–2012, he was a postdoctoral researcher at the Aerospace Engineering Department, Technion–Israel Institute of Technology, Haifa, Israel.

Currently, he works as a Professor at the Aerospace Engineering Department, Indian Institute of Science, Bangalore, India where he heads the Autonomous Vehicles Laboratory. His research focuses on guidance and control of autonomous vehicles.

He is an Associate Fellow of American Institute of Aeronautics and Astronautics (AIAA), a member of AIAA Guidance, Navigation, and Control Technical Committee, and a member of the editorial board for IEEE Transactions on Aerospace and Electronic Systems. During 2019–2024 he was an elected member of the Indian National Young Academy of Science (INYAS).
\end{IEEEbiography}

\end{document}